\documentclass[twocolumn]{autart}    

\usepackage{graphicx}          
\usepackage{amsmath, amssymb}   
\usepackage{mathtools}          
\usepackage{bm}                 
\usepackage{mathrsfs}           
\usepackage{algorithm}
\usepackage{algpseudocode}
\usepackage{tabularx}
\usepackage{booktabs}
\newtheorem{assumption}{Assumption}

\newtheorem{proposition}{Proposition}
\newtheorem{definition}{Definition}
\newtheorem{theorem}{Theorem}
\newtheorem{lemma}{Lemma}
\newenvironment{lemmastar}
{\par\medskip\noindent\textbf{Lemma:}\itshape\ }
{\par\medskip}
\newenvironment{theoremstar}
{\par\medskip\noindent\textbf{Theorem:}\itshape\ }
{\par\medskip}

\newcommand{\R}{\mathbb{R}}

\usepackage{subfig}
\usepackage{enumitem}

\usepackage{todonotes} 
\definecolor{pastel_mint_green}{RGB}{198, 240, 230} 
\definecolor{purple_custom}{RGB}{65,35,142}
\presetkeys{todonotes}{backgroundcolor=pastel_mint_green,linecolor=pastel_mint_green}{}

\newcounter{todocounter}

\let\oldtodo\todo  
\renewcommand{\todo}[1]{%
    \refstepcounter{todocounter}%
    \oldtodo[inline]{\thetodocounter \ - #1}%
}

\newcommand{\figref}[1]{Fig.~\ref{#1}}

\newcommand{\secref}[1]{Section~\ref{#1}}
\newcommand{\assref}[1]{Assumption~\ref{#1}}

\newcommand{\eqnref}[1]{equation~\eqref{#1}}
\newcommand{\lemref}[1]{Lemma~\eqref{#1}}

\usepackage{color,soul}

\usepackage[nolist]{acronym} 

\usepackage{comment}  

\newif\ifshowproofs
\showproofstrue        

\ifshowproofs
  \includecomment{proofver}
  \excludecomment{ideaver}
\else
  \excludecomment{proofver}
  \includecomment{ideaver}
\fi

\begin{document}

\begin{frontmatter}

\title{Unifying Hamilton-Jacobi Reachability and Reinforcement Learning}


\author[tudelft]{Prashant Solanki}\ead{p.solanki@tudelft.nl},
\author[tudelft]{Isabelle El-Hajj}\ead{i.z.el-hajj-1@tudelft.nl},        
\author[tudelft]{Jasper van Beers}\ead{j.j.vanbeers@tudelft.nl},  
\author[tudelft]{Erik-Jan van Kampen} \ead{e.vankampen@tudelft.nl},  
\author[tudelft]{Coen de Visser}\ead{c.c.devisser@tudelft.nl}

\address[tudelft]{Section of Control \& Simulation at the Faculty of Aerospace Engineering, Delft University of Technology, Kluyverweg 1, 2629HS, Delft, The Netherlands}             %
   

\begin{abstract}
We unify Hamilton-Jacobi (HJ) reachability and Reinforcement Learning (RL) through a proposed running cost formulation. We prove that the resultant travel-cost value function is the unique bounded viscosity solution of a time-dependent Hamilton-Jacobi Bellman (HJB) Partial Differential Equation (PDE) with zero terminal data, whose negative sublevel set equals the strict backward-reachable tube. Using a forward reparameterization and a contraction inducing Bellman update, we show that fixed points of small-step RL value iteration converge to the viscosity solution of the forward discounted HJB. Experiments on a classical benchmark validate this connection by demonstrating convergence of learned value functions toward semi-Lagrangian HJB solutions and by quantifying approximation error across the state space. These results empirically support the theoretical analysis, showing that the proposed framework preserves reachability-based safety semantics while remaining compatible with deep RL implementations.
\end{abstract}

\end{frontmatter}

\begin{acronym}[HJB] 
  \acro{brs}[BRS]{Backward Reachable Set}
  \acro{brt}[BRT]{Backward Reachable Tube}
  \acro{cbf}[CBF]{Control Barrier Function}
  \acro{cbvf}[CBVF]{Control Barrier Value Function}
  \acro{dpp}[DPP]{Dynamic Programming Principle}
  \acro{hj}[HJ]{Hamilton Jacobi}
  \acro{hjb}[HJB]{Hamilton Jacobi Bellman}
  \acro{hjvi}[HJVI]{Hamilton Jacobi Variational Inequality}
  \acro{mdp}[MDP]{Markov Decision Process}
  \acro{mdr}[MDR]{Minimum Discounted Reward}
  \acro{nn}[NN]{Neural Network}
  \acro{pde}[PDE]{Partial Differential Equation}
  \acro{qp}[QP]{Quadratic Programming}
  \acro{rl}[RL]{Reinforcement Learning}
  \acro{roi}[ROI]{Region of Interest}
  \acro{siren}[SIREN]{Sinusoidal Representation Network}
  \acro{td}[TD]{Temporal Difference}
\end{acronym}

\section{Introduction}

Safety is fundamental to the deployment of autonomous systems in uncertain and adversarial environments. From collision avoidance in air traffic management to motion planning for autonomous vehicles and safe learning for robots, the central challenge is to characterize the set of initial states from which trajectories can be kept out of failure regions over time. This safe set, equivalently the complement of the \ac{brs} or \ac{brt} of the unsafe set, underpins formal verification, supervisory control, and online safety filtering. \ac{hj} reachability provides a rigorous framework for such analysis by formulating safety as a differential game whose value function solves a Hamilton Jacobi (\ac{hj}) \ac{pde} or \ac{hjvi} \cite{bansal2017hj,lygeros2004reachability,mitchell2005time}.

Despite its broad impact in many safety-critical domains, classical \ac{hj} solvers suffer from the curse of dimensionality. The computational burden of gridding grows exponentially with state dimension, often rendering direct solutions impractical beyond roughly six dimensions \cite{akametalu2014reachability,bansal2017hj,chen2018decomposition,darbon2016algorithms,mitchell2005time}. To mitigate this, decomposition methods exploit separability \cite{chen2018decomposition}, neural approximators such as DeepReach \cite{bansal2021deepreach}, convex relaxations \cite{yin2021backward}, and operator-theoretic approaches including Hopf and Koopman methods based formulations \cite{lee2020hopf,umathe2022reachability} offer approximate alternatives. Relatedly, control barrier functions (CBFs) provide real time certificates of forward invariance through \ac{qp} based controllers \cite{ames2019control}, while hybrid constructions such as \ac{cbvf} combine barrier ideas with discounted \ac{hj} value functions \cite{choi2021robust}. These methods, however, typically presume accurate models, may be conservative, and can still incur substantial offline computation.

\ac{rl} offers a complementary, data-driven paradigm for long-horizon decision making and has demonstrated strong empirical scalability on high-dimensional and nonlinear control problems~\cite{mnih2015human,sutton1998reinforcement}. However, standard \ac{rl} is built around additive cumulative, often discounted, return, whereas classical \ac{hj} reachability is typically expressed through minimum- or maximum-over-time criteria. As a result, classical temporal-difference updates do not directly encode the usual \ac{hj} reachability objective. Moreover, in the classical \ac{hj} setting, the Bellman operator is undiscounted and therefore ceases to be contractive~\cite{fisac2019bridging}, removing the convergence guarantees that underpin much of standard \ac{rl} theory. Consequently, pure \ac{rl} methods do not directly inherit the rigorous safety and robustness guarantees associated with Hamilton Jacobi reachability.

A growing body of work explores the interface between \ac{hj} reachability and \ac{rl}. Some approaches inject reachability-based structure into learning, for example by using precomputed reachable sets to guide exploration or impose safety filters \cite{nagami2021hjb}, by reinterpreting policy iteration through a \ac{pde} lens \cite{wiltzer2022distributional}, or by deriving actor critic schemes from continuous-time \ac{hjb} equations \cite{ganai2023iterative}. Others use \ac{hj} solutions to shape rewards or initialize policies \cite{chen2023hjrl,bansal2017hj}. A related but distinct line of work introduces reachability quantities into constrained reinforcement learning. In particular, \cite{yu2022reachability} formulate reachability-constrained reinforcement learning through a safety value function enforced via a statewise Lagrangian constrained optimization, rather than deriving a direct Bellman/\ac{hjb} equivalence for a single discounted value function.

Recent discounted Hamilton Jacobi formulations are also relevant. Xue et al.~\cite{xue2022differential} study discounted differential games for robust controlled invariant sets, characterizing lower and upper invariant sets as zero level sets of viscosity solutions, whereas Li et al.~\cite{li2026converse} study discrete time infinite horizon reach-avoid learning with a Lipschitz continuous, contractive value function and post-learning certification. Among discounted reachability-learning formulations, Fisac et al.~\cite{fisac2019bridging} obtain contraction through a probabilistic discounting interpretation, but do not establish exact Hamilton Jacobi reachability semantics or a viscosity characterization. Akametalu et al.~\cite{akametalu2023minimum} obtain a discounted \ac{hjvi} and contraction through MDR, but exact agreement with the true reachable set is recovered only in the zero-discount limit. While these works are closely related, they are based on min/max-over-time safety values, constrained formulations, or discounted reachability surrogates rather than a single continuous time cumulative running cost objective. To summarize, many existing approaches either (i) preserve \ac{hjb} semantics at the cost of compatibility with standard \ac{rl} routines or (ii) impose discounting on the min/max-over-time reachability objective for \ac{rl} compatibility but sacrifice \ac{hjb} semantics in doing so (meaning that the learned value functions do not accurately represent the true safe sets). None have yet managed to satisfy both the preservation of \ac {hj} reachability semantics while retaining compatibility with standard \ac{rl} routines. 

\begin{proofver}
\textbf{In this paper}, we develop such a unified value-function formalism that rigorously connects \ac{rl} and \ac{hj} reachability through a travel-cost construction. The key observation is that, after the usual sign change between cost minimization and reward maximization, the discounted travel-cost objective is already an additive cumulative discounted return of the type used in standard \ac{rl}. This makes the connection to \ac{rl} direct: rather than approximating a min/max-over-time reachability objective or imposing reachability through an external constraint, we work with a single discounted value function whose sign encodes strict reachability and whose Bellman operator is contractive.

Our formulation differs from \cite{akametalu2023minimum,fisac2019bridging,yu2022reachability} in two essential ways. First, we show that a calibrated running cost (i.e. zero off target and strictly negative on target) recovers strict \ac{brt} semantics without terminal penalties. Second, under relative exponential discounting, we derive a discounted \ac{dpp} and an equivalent forward discounted \ac{hjb} whose exact one-step Bellman operator has the value function $W_\lambda$ as its fixed point. Moreover, consistent time discretizations of the dynamics and running cost yield monotone, stable, and consistent Bellman schemes whose fixed points converge to the viscosity solution in the small-step limit via Barles Souganidis theory \cite{barles1991convergence}. In this way, our framework provides a direct bridge between continuous-time \ac{hjb} analysis and \ac{rl}-style Bellman updates while preserving exact reachability semantics. This allows us to solve reachability problems using standard \ac{rl} routines. 
\end{proofver}

\begin{ideaver}
\textbf{In this paper}, we develop such a unified value-function formalism that rigorously connects \ac{rl} and \ac{hj} reachability through a travel-cost construction. The key observation is that, after the usual sign change between cost minimization and reward maximization, the discounted travel-cost objective is already an additive cumulative discounted return of the type used in standard \ac{rl}. This makes the connection to \ac{rl} direct. Such a formulation differs from \cite{akametalu2023minimum,fisac2019bridging,yu2022reachability} in two essential ways. First, we show that a calibrated running cost (i.e. zero off target and strictly negative on target) recovers strict \ac{brt} semantics without terminal penalties. Second, under relative exponential discounting, we derive a discounted \ac{dpp} and an equivalent forward discounted \ac{hjb} whose exact one-step Bellman operator has the value function $W_\lambda$ as its fixed point. 
\end{ideaver}

\subsection*{Contributions}
\begin{itemize}
\item \textbf{Travel-cost \ac{hjb} with exact reachability semantics.} We define a running-cost value function that is a viscosity solution of a time-dependent \ac{hjb} and prove that the strict backward reachable tube is exactly its negative sublevel set, while its complement is the zero level set. This yields exact reachability semantics without terminal penalties \cite{mitchell2005time,bansal2017hj,chen2023hjrl}.

\item \textbf{Relative discount, \ac{dpp}, and contraction.} For weights $e^{\lambda(t-s)}$, we derive a discounted \ac{dpp} in which the continuation term is multiplied by $e^{-\lambda\sigma}$, prove that the induced Bellman operator is a strict contraction for $\lambda>0$, and establish boundedness, spatial Lipschitz continuity, and time continuity of the discounted value while preserving reachability semantics.

\item \textbf{Forward discounted \ac{hjb} $\leftrightarrow$ Bellman equivalence.} Using a forward reparameterization, we show that exact one-step Bellman fixed points recover $W_\lambda$, and that consistent discretized Bellman schemes converge to the forward \ac{hjb} viscosity solution as the step shrinks. In particular, the scheme is monotone, stable, and consistent, so its fixed points converge by Barles Souganidis theory \cite{barles1991convergence}. We also derive a residual identity linking small-step Bellman and \ac{hjb} residuals, clarifying why vanishing Bellman residual enforces vanishing \ac{hjb} residual in the small-step limit \cite{barles1991convergence,falcone2013semi}.

\item \textbf{\ac{rl}-compatible, safety-aware value learning.} Because the discounted travel-cost objective is already a cumulative discounted return, the framework provides a principled route to sample-based safety-aware value learning and policy optimization that remains aligned with continuous-time optimal control while preserving \ac{hj}-level reachability semantics, complementing prior heuristic or problem-specific bridges \cite{fisac2019bridging,akametalu2014reachability,chen2023hjrl}.
\end{itemize}

\begin{rem}[Scope: reach vs.\ avoid]
Owing to space constraints, we restrict attention to the reach formulation. The avoid formulation is entirely analogous: it is obtained by replacing the minimizing control (infimum) in the Bellman/\ac{hjb} operator with a maximizing one (supremum). Concretely, if the reach operator reads
\[
(\mathcal{T}V)(x)=\inf_{u\in\mathcal U}\Big\{h(x,u)+\gamma\,V\big(f(x,u)\big)\Big\},
\]
then the avoid operator is
\[
(\mathcal{T}_{\mathrm{avoid}}V)(x)=\sup_{u\in\mathcal U}\Big\{h(x,u)+\gamma\,V\big(f(x,u)\big)\Big\}.
\]
All statements and proofs carry over after replacing the minimizing control with a maximizing one.
\end{rem}

\section{Problem Setup}\label{sec:problem_statement}

This section establishes the notation and standing assumptions for a finite-horizon reachability problem. We define the associated cost/value functionals that will be used throughout, providing the problem statement and repository of assumptions for the \ac{dpp}/\ac{hjb} analysis in \secref{sec:hjb} and \secref{sec:discounted_hjb}.

\subsection{System Dynamics}
We consider a continuous-time deterministic control system governed by
\begin{equation}\label{eq-main-ODE}
    \dot{x}(s) = f(x(s), u(s)), \qquad x(t) = x \in \mathbb{R}^n, \qquad s \in [t, T],
\end{equation}
where $x(s) \in \mathbb{R}^n$ is the state trajectory and $u(s) \in \mathcal{U} \subset \mathbb{R}^m$ is the control input. We define $\mathcal{M}(t)$ as the set of all control policies applicable at time $t$:
\[
    \mathcal{M}(t)\equiv\{u:[t,T]\rightarrow \mathcal{U}\mid u \text{ measurable} \}.
\]

\begin{rem}
The numerical studies in Section~\ref{sec:method} are carried out on compact regions of interest (ROIs), and the physical dynamics used there need only be well-defined and locally Lipschitz on an open neighborhood of those compact sets. For the Hamilton Jacobi analysis in Sections~\ref{sec:hjb} and \ref{sec:discounted_hjb}, we use the same symbol $f$ to denote a globally defined auxiliary vector field that coincides with the physical dynamics on an open neighborhood of the ROIs reported in the experiments. All regularity assumptions below are imposed on this globally defined auxiliary field. Since the auxiliary and physical dynamics agree on the ROIs used in the experiments, the theoretical and physical trajectories coincide throughout the domains relevant to the numerical study. Such an extension-based reformulation is standard in Hamilton Jacobi analysis on bounded regions \cite{li2026converse,xue2019inner}
\end{rem}

This setup covers the experimentally relevant case of locally Lipschitz physical dynamics on bounded domains while still allowing the \ac{hjb} theory to be stated on all of $\mathbb R^n$. The globally Lipschitz physical-dynamics setting is recovered as the special case in which the auxiliary field equals the physical dynamics on all of $\mathbb R^n$.

In this paper, we assume that the globally defined vector field $f$ in \eqnref{eq-main-ODE} satisfies the following assumptions.

\begin{assumption}[$\mathcal{U}$ is compact]\label{a:U_compact}
    Let \( \mathcal{U} \subset \mathbb{R}^m \). We assume that \( \mathcal{U} \) is compact, i.e., \( \mathcal{U} \) is closed and bounded.
\end{assumption}

\begin{assumption}[Global Lipschitz continuity in $x$]\label{a:f_lipschitz}
There exists $L_f > 0$ such that
\begin{multline}
    \| f(x_1, u) - f(x_2, u) \| \leq L_f \| x_1 - x_2 \|, \\
    \forall x_1, x_2 \in \mathbb{R}^n, \, u \in \mathcal{U}.
\end{multline}
\end{assumption}

\begin{assumption}[Linear growth]\label{a:f_uniform}
There exists $C_f > 0$ such that
\[
\| f(x, u) \| \leq C_f(1+\|x\|), \qquad \forall x \in \mathbb{R}^n, \, u \in \mathcal{U}.
\]
\end{assumption}

\begin{assumption}[Continuity in $u$ for $f$]\label{a:f_continuity_u}
For each $x \in \mathbb{R}^n$, the map
\[
    u \mapsto f(x,u)
\]
is continuous on $\mathcal{U}$.
\end{assumption}

\begin{rem}[On Assumption~\ref{a:f_uniform}]
Assumption~\ref{a:f_uniform} is stated explicitly because it is convenient in later boundedness and time-continuity estimates. It is automatically satisfied, for example, when the globally defined auxiliary field $f$ is globally Lipschitz in $x$ and continuous in $u$ on the compact control set $\mathcal U$. This extension based viewpoint is standard in Hamilton Jacobi analysis when the physical dynamics are only locally well behaved on the compact region relevant to verification or computation \cite{li2026converse,xue2019inner}
\end{rem}

Under Assumptions~\ref{a:U_compact}--\ref{a:f_continuity_u}, for any initial condition $(t,x)$ and any admissible control $u \in \mathcal M(t)$, the state equation \eqnref{eq-main-ODE} admits a unique Carath\'eodory trajectory on $[t,T]$. We denote this trajectory by
\[
    x^u_{t,x}(\cdot):[t,T]\to\mathbb R^n.
\]

\vspace{0.5em}

We define a uniformly continuous \textit{travel cost} function
\[
h : [0, T] \times \mathbb{R}^n \times \mathcal{U} \to \mathbb{R},
\]
and we make the following assumptions regarding this travel cost function.

\begin{assumption}[Lipschitz continuity in $x$]\label{a:h_lipschitz}
There exists $L_h > 0$ such that
\begin{multline}
    | h(s, x_1, u) - h(s, x_2, u) | \leq L_h \| x_1 - x_2 \|, \\
    \forall x_1, x_2 \in \mathbb{R}^n, \, u \in \mathcal{U}, \, s \in [0, T].
\end{multline}
\end{assumption}

\begin{assumption}[Uniform boundedness]\label{a:h_uniform}
There exists $M_h > 0$ such that
\[
| h(s, x, u) | \leq M_h, \qquad \forall (s, x, u) \in [0, T] \times \mathbb{R}^n \times \mathcal{U}.
\]
\end{assumption}

\begin{assumption}[Continuity in $u$ for $h$]\label{a:h_continuity_u}
For each $(s,x) \in [0,T]\times \mathbb{R}^n$, the map
\[
    u \mapsto h(s,x,u)
\]
is continuous on $\mathcal{U}$.
\end{assumption}

\subsection{Travel-Cost Value Function}
We define the payoff associated with \eqnref{eq-main-ODE} by
\begin{equation}\label{eq_main_payoff_function}
P(t,x,u) = \int_{t}^{T} h\big(s,x^u_{t,x}(s),u(s)\big)\,ds,
\end{equation}
which the control policy $u(\cdot)$ seeks to minimize.

The corresponding value function is
\begin{equation}\label{eq-main-value-func}
    V(t,x) = \inf_{u \in \mathcal{M}(t)} P(t,x,u).
\end{equation}

For an initial condition $(t,x)$ and an admissible control $u \in \mathcal{M}(t)$, we denote by
\[
    x^u_{t,x}(\cdot) : [t,T] \to \mathbb{R}^n
\]
the unique trajectory solving \eqnref{eq-main-ODE}. For each $s \in [t,T]$, the notation
\[
    x^u_{t,x}(s) \in \mathbb{R}^n
\]
denotes the state at time $s$ of the trajectory of \(\dot{x}=f(x,u)\) initialized at \(x\) at time \(t\) and driven by the control \(u(\cdot)\). When \(t\) and the control law are clear from context, we abbreviate the trajectory to \(x(\cdot)\) and the state at time \(s\) to \(x(s)\).
\section{\ac{hjb} PDE for the Travel-Cost Value Function}\label{sec:hjb}

In this section, we work under the standing regularity assumptions on the globally defined auxiliary dynamics introduced in Section~\ref{sec:problem_statement}, and encode the open target $\mathcal{T}\subset\mathbb{R}^n$ through a calibrated running cost that vanishes off target and is strictly negative on target.

Under these assumptions, the resulting value function is the unique bounded viscosity solution of the time dependent \ac{hjb} equation with zero terminal data. Its sign exactly recovers strict backward reachability: the negative sublevel set $\{V(t,\cdot)<0\}$ coincides with the strict \ac{brt}, whereas $\{V(t,\cdot)=0\}$ characterizes states from which the target can be avoided almost everywhere in time.

\begin{theorem}[\ac{hjb} characterization; viscosity sense]\label{thm:hjb}
For $(t,x)\in[0,T]\times\mathbb R^n$, let
\begin{equation}
    \begin{aligned}
       &V(t,x):=\inf_{u(\cdot)\in\mathcal M(t)} \int_{t}^{T} h\big(s,x^{u}_{t,x}(s),u(s)\big)\,ds,\\
       &V(T,x)= 0, 
    \end{aligned}
    \label{eq:V_undisc_strict}
\end{equation}
and define
\begin{equation}\label{eq:H_def_thm}
H(t,x,p):=\inf_{u\in\mathcal U}\big\{\,h(t,x,u)+p\!\cdot f(x,u)\,\big\}
\end{equation}
Under the standing assumptions,
$V$ is a unique and bounded viscosity solution of
\begin{equation}\label{eq:HJB_PDE}
    \begin{aligned}
        &V_t(t,x)+H\!\big(t,x,\nabla_x V(t,x)\big)=0
        \quad\text{on }[0,T)\times\mathbb R^n, \\
        &V(T,x)=0.
    \end{aligned}
\end{equation}
\end{theorem}

\begin{pf}
A complete, proof of the \ac{hjb} characterization is a direct specialization of standard result in \cite{evans1984differential}.
\end{pf}

\subsection{Reachability via Running Cost (Strict BRT)}
\label{subsec:reach_strict}
We now interpret the sign of $V(t,x)$ in terms of backward reachability.

\textbf{Sign/calibration of running/travel cost.}
We impose
\begin{align}
\text{(S0)}\quad & h(s,x,u)=0 
&& \forall\,s\in[0,T],\ \forall\,u\in\mathcal U,\ \forall\,x\notin\mathcal T,
\label{ass:S0}\\
\text{(S1)}\quad & \inf_{u\in\mathcal U} h(s,x,u)<0 
&& \forall\,s\in[0,T],\ \forall\,x\in\mathcal T
\label{ass:S1}
\end{align}

\textbf{Strict \ac{brt}.} For $V(t,x)$ defined as in \eqnref{eq:V_undisc_strict}, the strict \ac{brt} is defined as follows.
\begin{equation}
\mathcal R(t) := \Big\{\,x:\ \exists\,u(\cdot),\ \exists\,s\in[t,T)\ \text{s.t.}\ 
x^u_{t,x}(s)\in\mathcal T \Big\}
\label{eq:Rstrict_def}
\end{equation}

\begin{proposition}[Negative sublevel implies \ac{brt}]
\label{prop:neg_equals_strictBRT}
Under \eqnref{ass:S0}--\eqnref{ass:S1}, for every $t\in[0,T)$,
\begin{equation}\label{eq:strict_equiv_neg}
\mathcal R(t)\;=\;\{\,x:\ V(t,x)<0\,\}.
\end{equation}
\end{proposition}

\begin{pf}
Soundness ($\{x: V(t, x)<0\}\subseteq\mathcal R(t)$).\\
If a trajectory stays off $\mathcal T$ on $[t,T)$, then by \eqnref{ass:S0}
the integrand is $0$ almost everywhere (a.e). Hence, its integral is $0$. Minimizing gives
$V(t,x)\ge 0$. Thus $V(t,x)<0$ implies a hit of $\mathcal T$ at some $s<T$.

Completeness ($\mathcal R(t)\subseteq\{x: V(t, x)<0\}$).\\
Fix $x\in\mathcal R(t)$. Then $\exists\,u_0(\cdot)$ and
$s_0\in[t,T)$ with $x^{u_0}_{t,x}(s_0)\in\mathcal T$. Since $\mathcal T$ is open,
pick $\rho>0$ with $B_\rho\!\big(x^{u_0}_{t,x}(s_0)\big)\subset\mathcal T$.
By \eqnref{ass:S1} and uniform continuity of $h$, there exist
$u^-\in\mathcal U$, $\eta>0$, and $\delta>0$ such that
\begin{equation}\label{eq:neg_pulse}
h(s,y,u^-)\ \le\ -\eta
\quad \forall\,s\in[s_0,s_0+\delta],\ \forall\,y\in B_\rho\!\big(x^{u_0}_{t,x}(s_0)\big)
\end{equation}
By continuity of trajectories, holding the constant control $u^-$ from $s_0$
keeps the state in $B_\rho$ on $[s_0,s_0+\delta']$ for some
$0<\delta'\le \min\{\delta,\,T-s_0\}$. Define the concatenated control
\begin{equation}\label{eq:concat_u_strict}
u^\ast(s)=
\begin{cases}
u_0(s), & s\in[t,s_0),\\
u^-, & s\in[s_0,s_0+\delta'],\\
\text{arbitrary}, & s\in[s_0+\delta',T].
\end{cases}
\end{equation}
By \eqnref{ass:S0}, the off-target cost (on $[t,s_0)$ and whenever the trajectory
exits $\mathcal T$) is identically $0$. Over $[s_0,s_0+\delta']\subset[t,T)$,
\eqnref{eq:neg_pulse} gives
\begin{equation}\label{eq:neg_total_strict}
\int_{t}^{T}\! h\big(s,x^{u^\ast}(s),u^\ast(s)\big)\,ds
\ \le\ \int_{s_0}^{s_0+\delta'}\! (-\eta)\,ds
\ =\ -\eta\,\delta' \ <\ 0.
\end{equation}
Hence $V(t,x)\le -\eta\,\delta'<0$.
\end{pf}

\begin{proposition}[Zero level equals complement]
\label{prop:zero_equals_complement}
Under \eqnref{ass:S0}--\eqnref{ass:S1}, for every $t\in[0,T)$,
\begin{equation}\label{eq:zero_equals_comp}
\big(\mathcal R(t)\big)^{\complement}\;=\;\{\,x:\ V(t,x)=0\,\}.
\end{equation}
\end{proposition}

\begin{pf}
If $x\notin\mathcal R(t)$, then for every admissible control $u(\cdot)\in\mathcal M(t)$
the corresponding trajectory satisfies $x^u_{t,x}(s)\notin\mathcal T$ for all $s\in[t,T)$.
By \eqref{ass:S0} we have $h(s,x^u_{t,x}(s),u(s))=0$ for a.e.\ $s\in[t,T)$, hence
\[
\int_t^T h\big(s,x^u_{t,x}(s),u(s)\big)\,ds = 0 \qquad \forall\,u(\cdot)\in\mathcal M(t).
\]
Taking the infimum over $u(\cdot)$ yields $V(t,x)=0$.
Conversely, if $x\in\mathcal R(t)$, then Proposition~\ref{prop:neg_equals_strictBRT} implies
$V(t,x)<0$, hence $x\notin\{V(t,\cdot)=0\}$. Therefore
$\{x:V(t,x)=0\}=(\mathcal R(t))^{\complement}$.
\end{pf}



\begin{rem}
    The purpose of first presenting the undiscounted case is to isolate the strict-reachability sign semantics of the calibrated running-cost formulation. The discounted case considered next preserves these semantics while adding the contraction property needed for the RL/Bellman connection.
\end{rem}
\section{Relative Exponential Discount}\label{sec:discounted_hjb}
Section \ref{sec:hjb} established that a running cost value function calibrated to be identically zero off the (open) target and strictly negative on it, solves a time‐dependent \ac{hjb} and exactly encodes strict backward reachability: the strict \ac{brt} is the negative sublevel set of $V(t,\cdot)$, while its complement is the zero level. In this section, we retain these reachability semantics but introduce a relative exponential discount, weighting the integrand by $e^{\lambda(t-s)}$. Because the weights are positive, the sign logic underlying strict capture is preserved, so the same sublevel/zero-level characterization of the \ac{brt} holds. At the same time, the \ac{dpp} acquires a factor $e^{-\lambda\sigma}$ on the continuation term, yielding a strictly contractive one step Bellman operator for $\lambda>0$, and the PDE gains the stabilizing zeroth order term $-\lambda V$. This discounted formulation will be pivotal later: under a forward reparametrization it aligns exactly with the $\gamma=e^{-\lambda\sigma}$ discounted Bellman update used in RL, enabling both convergence guarantees and a clean bridge between \ac{hj} reachability and reinforcement learning. \\

\begin{rem}
    \textbf{Why this formulation is needed.}
The strict contraction obtained below is induced by the relative exponential discount, which introduces the factor \(e^{-\lambda\sigma}\) on the continuation term in the discounted DPP. However, discounting alone is not the main point: the running-cost formulation is introduced so that this contraction is obtained without sacrificing exact strict-reachability semantics. In classical reachability formulations based on minimum/maximum-over-time criteria, the Bellman recursion is obstacle-type and is generally nonexpansive rather than strictly contractive. If discounting is applied directly to such formulations, one may recover contraction, but the resulting value need not preserve the exact Hamilton--Jacobi reachable set for finite discount, since late events become downweighted. By contrast, in the present travel-cost formulation the calibrated sign structure of \(h\) yields exact strict-reachability semantics, while the positive relative discount preserves that sign structure and at the same time yields a strict Bellman contraction.
\end{rem}

\textbf{Discounted problem}
Fix $\lambda\in\mathbb R^{+}$. For $(t,x)\in[0,T]\times\mathbb R^n$ and $u\in\mathcal M(t)$ define
\begin{align}
J_\lambda(t,x;u)
&:= \int_{s=t}^{T} e^{\lambda(t-s)}\,
      h\!\big(s,\,x^u_{t,x}(s),\,u(s)\big)\,ds,
\label{eq:Jlambda_def}
\\[-1pt]
V_\lambda(t,x)
&:= \inf_{u\in\mathcal M(t)} J_\lambda(t,x;u),
\qquad V_\lambda(T,x)=0.
\label{eq:Vlambda_def}
\end{align}
Under Assumptions~\ref{a:f_lipschitz}, \ref{a:f_continuity_u} and measurability of $u$, the trajectory $s\mapsto x^u_{t,x}(s)$ exists uniquely on $[t,T]$ in the Carath\'eodory sense. Based on \assref{a:h_uniform} and \assref{a:h_continuity_u}, the map $s\mapsto h\!\big(s,x^u_{t,x}(s),u(s)\big)$ is measurable and bounded, hence integrable.

\begin{proofver}
\begin{lemma}[Well-posedness]\label{lem:wellposed}
Under \assref{a:h_uniform},
\begin{multline}
|J_\lambda(t,x;u)|
\le M_h\!\int_t^T e^{\lambda(t-s)}ds \\
 =\begin{cases}
\frac{M_h}{\lambda}\big(1-e^{-\lambda(T-t)}\big), & \lambda>0,\\[4pt]
M_h\,(T-t), & \lambda=0,
\end{cases}
\label{eq:J_bound_wp}
\end{multline}
for all $u\in\mathcal M(t)$. In particular $J_\lambda(t,x;u)\in\mathbb R$ and $V_\lambda(t,x)\in\mathbb R$.
\end{lemma}
\begin{pf}
Immediate from $|h|\le M_h$ and \eqnref{eq:Jlambda_def}.
\end{pf}
\end{proofver}


We first establish a discounted \ac{dpp} for $V_\lambda$, which splits the objective into a
short-horizon running cost and a discounted continuation value. This identity is the main
tool used to derive the HJB characterization.

\begin{lemma}[\ac{dpp} with relative discount]\label{lem:DPP_rel_fnl}
For any $(t,x)\in[0,T]\times\mathbb R^n$ and $\sigma\in[0,\,T-t]$,
\begin{align}
V_\lambda(t,x)
&= \inf_{u\in\mathcal M(t)}\Bigg\{
\int_{t}^{t+\sigma}\! e^{\lambda(t-s)}\,
   h\!\big(s,\,x^u_{t,x}(s),\,u(s)\big)\,ds
\notag\\[-1pt]&\hspace{7.45em}
+\,e^{-\lambda\sigma}\,
   V_\lambda\!\big(t+\sigma,\,x^u_{t,x}(t+\sigma)\big)
\Bigg\}.
\label{eq:DPP_rel_fnl_pap}
\end{align}
\end{lemma}

\begin{pf}
The proof is given in Appendix~\ref{lem:DPP_rel_fnl_appen}.
\end{pf}

\begin{proofver}
We next show that $V_\lambda$ is uniformly bounded. This guarantees well-posedness (and, for
$\lambda>0$, the infinite-horizon case) and provides a global constant used in later estimates.

\begin{lemma}[Boundedness]\label{lem:bounded}
Under \assref{a:h_uniform},
\begin{multline}
|V_\lambda(t,x)|\ \le\ 
\int_{0}^{T-t}\! e^{-\lambda r}M_h\,dr
= \\ \begin{cases}
\frac{M_h}{\lambda}\big(1-e^{-\lambda(T-t)}\big), & \lambda>0,\\[4pt]
M_h\,(T-t), & \lambda=0.
\end{cases}
\label{eq:bound_value_pap}
\end{multline}
\end{lemma}
\begin{pf}
The proof is given in Appendix~\ref{lem:bounded_appen}.
\end{pf}



We show $V_\lambda(t,\cdot)$ is Lipschitz in $x$ to obtain the spatial regularity needed for
continuity of $V_\lambda$ and for the comparison/uniqueness argument.

\begin{lemma}[Lipschitz in state]\label{lem:Lip_x}
Given the above assumptions, for fixed $t$,
\begin{equation}
|V_\lambda(t,x_1)-V_\lambda(t,x_2)|
\ \le\ \Gamma_\lambda(t)\,\|x_1-x_2\|,
\label{eq:Lip_bound_V_pap}
\end{equation}
where
\[\Gamma_\lambda(t):=L_h\int_0^{T-t} e^{(L_f-\lambda)r}\,dr,\]
\end{lemma}
\begin{pf}
The proof is given in Appendix~\ref{lem:Lip_x_appen}.
\end{pf}


Next establish continuity in $t$ so that $V_\lambda$ is continuous on $[0,T]\times\R^n$, which
is a standing requirement for the viscosity framework and the uniqueness result.

\begin{lemma}[Time continuity]\label{lem:time_cont}
Under \assref{a:h_uniform}, \assref{a:f_uniform}, and \lemref{lem:Lip_x}, for $\sigma\in[0,T-t]$,
\begin{multline}\label{eq:time_cont_pap}
    |V_\lambda(t+\sigma,x)-V_\lambda(t,x)|
\ \le\ M_h\!\int_{0}^{\sigma}\! e^{-\lambda r}dr\\
+\,e^{-\lambda\sigma}\Gamma_\lambda(t+\sigma)\big(e^{C_f\sigma}-1\big)(\|x\|+1)
\, \\ +\,|1-e^{-\lambda\sigma}|\,\|V_\lambda\|_\infty
\end{multline}.
\end{lemma}

\begin{pf}
The proof is given in Appendix~\ref{lem:time_cont_appen}.
\end{pf}


Let us define the following. 
\begin{equation}
H(t,x,p):=\inf_{u\in\mathcal U}
\big\{\,h(t,x,u)+p\!\cdot f(x,u)\,\big\},
\label{eq:H_def_time}
\end{equation}
and, for $\phi\in C^1$, set
\begin{multline}
\Lambda_\lambda(s,x,u;\phi)
:= \phi_t(s,x)+D_x\phi(s,x)\!\cdot f(x,u) \\
 + h(s,x,u) - \lambda\,\phi(s,x).
\label{eq:Lambda_def}
\end{multline}


Finally, we combine the \ac{dpp} with rest of the lemmas to prove that $V_\lambda$ is the
(unique) bounded continuous viscosity solution of the discounted HJB equation.
\end{proofver}
\begin{ideaver}
Now, we prove that $V_\lambda$ is the
(unique) bounded continuous viscosity solution of the discounted HJB equation. We will only show the viscosity solution proof, for the proof of boundedness and Lipchitz continuity please refer to the arXiv version of this paper \cite{solanki2026formalizingrelationshiphamiltonjacobireachability}.
\end{ideaver}


\begin{theorem}[Viscosity characterization]\label{thm:visc_V}
Under \assref{a:U_compact}-\assref{a:h_continuity_u},
$V_\lambda$ is a bounded, continuous and unique viscosity solution of
\begin{multline}
V_{\lambda,t}(t,x)+H\!\big(t,x,\nabla_x V_\lambda(t,x)\big)
-\lambda\,V_\lambda(t,x)=0, \\
V_\lambda(T,x)=0
\label{eq:HJB_rel_pap}
\end{multline}
\end{theorem}

\begin{pf}


The proof is given in Appendix~\ref{thm:visc_V_appen}.
\end{pf}



\subsection{Reachability Encoding with Relative Discount (Strict \ac{brt})}
\label{subsec:reach_rel_discount}
We encode strict backward reachability as a discounted optimal-control problem with a
relative exponential weight $\omega_t(s)=e^{\lambda(t-s)}$ and a sign-calibrated running cost
$h$ that is identically zero outside the target and strictly negative inside
(\eqnref{eq:S0_rel}-\eqnref{eq:S2_rel}). Under the standing regularity, the associated value
$V_\lambda$ solves the discounted \ac{hjb} and its negative sublevel set recovers exactly the strict \ac{brt}
(\eqnref{prop:BRT_negative_rel_strict}), while the zero level set matches its complement
(\eqnref{prop:BRT_zero_rel_strict}); the statement extends to infinite horizon when $\lambda>0$.

\textbf{Standing regularity.}
Consider \assref{a:U_compact}, \assref{a:f_lipschitz}, \assref{a:h_lipschitz}, \assref{a:h_uniform}, and \assref{a:h_continuity_u}. Moreover, let the target $\mathcal T\subset\mathbb R^n$ be open. Fix $\lambda\ge 0$ and define
\begin{equation}\label{eq:omega_rel}
\omega_t(s):=e^{\lambda(t-s)}\in(0,\infty),\qquad s\in[t,T].
\end{equation}

\textbf{Value function and strict \ac{brt}.} For $(t,x)\in[0,T]\times\R^n$,
\begin{multline}\label{eq:Vlambda_reach_strict}
    V_\lambda(t,x)
:= \inf_{u(\cdot)\in\mathcal M(t)}
\int_{t}^{T}\! \omega_t(s)\,
   h\!\big(s,\,x^{u}_{t,x}(s),\,u(s)\big)\,ds,
\\ V_\lambda(T,x)=0,
\end{multline}
\begin{multline}
\mathcal R(t)
:= \Big\{\,x\in\R^n:\ \exists\,u(\cdot)\in\mathcal M(t),\ 
\exists\,s\in[t,T)\ \\ \text{s.t. } x^{u}_{t,x}(s)\in\mathcal T \Big\}.
\label{eq:BRT_def_rel_strict}
\end{multline}
\textbf{Sign/Calibration (relative).}
\begin{align}
\text{(S0$_\lambda$)}\quad & h(s,x,u)=0,
&&\forall\,x\notin\mathcal T,\ \forall\,(s,u),
\label{eq:S0_rel}\\
\text{(S2$_\lambda$)}\quad & \inf_{u\in\mathcal U} h(s,x,u)<0,
&&\forall\,x\in\mathcal T,\ \forall\,s\in[0,T].
\label{eq:S2_rel}
\end{align}
By compactness of $\mathcal U$ and continuity in $u$, the infimum in \eqnref{eq:S2_rel} is attained.
If $h$ is continuous in $(s,x)$, \eqnref{eq:S2_rel} yields uniform negativity on a small neighborhood of each $(s,x)\in[0,T]\times\mathcal T$.

\begin{proposition}[Negative sublevel equals strict \ac{brt}]
\label{prop:BRT_negative_rel_strict}
Under \eqnref{eq:S2_rel} and the standing regularity,
for every $t\in[0,T)$,
\begin{equation}\label{eq:BRT_negative_rel_strict}
\mathcal R(t)\ =\ \{\,x\in\R^n:\ V_\lambda(t,x)<0\,\}.
\end{equation}
The statement also holds for the infinite horizon $T=\infty$ when $\lambda>0$.
\end{proposition}
\begin{pf}
The argument is the same as in Proposition~\ref{prop:neg_equals_strictBRT}. The only difference
is the multiplicative discount factor. Since for all $s\in[t,T]$ we have
$e^{\lambda(t-s)}>0$, multiplying $h$ by $e^{\lambda(t-s)}$ cannot change the sign of any
negative (or zero) contribution. Thus the proof follows same logic. 
\end{pf}


\begin{proposition}[Zero level equals complement]
\label{prop:BRT_zero_rel_strict}
Under \eqnref{eq:S0_rel}-\eqnref{eq:S2_rel}, for every $t\in[0,T)$,
\begin{equation}\label{eq:BRT_zero_rel_strict}
\big(\mathcal R(t)\big)^{\complement}\ =\ \{\,x\in\R^n:\ V_\lambda(t,x)=0\,\}.
\end{equation}
The same holds for $T=\infty$ when $\lambda>0$.
\end{proposition}
\begin{pf}
The proof is identical to Proposition~\ref{prop:zero_equals_complement}, with $V$ replaced by
$V_\lambda$. This is due to that fact that $e^{\lambda(t-s)}>0$, multiplying $h$ by $e^{\lambda(t-s)}$ cannot change the sign of any negative (or zero) contribution. Thus the proof follows same logic.
\end{pf}

\begin{rem}[Endpoint $T$ and strictness]
Integrals are taken over $[t,T]$, while reachability uses $[t,T)$. Since $\{T\}$ has measure zero,
including $T$ in \eqnref{eq:Vlambda_reach_strict} does not affect $V_\lambda$, and the strict tube
in \eqnref{eq:BRT_def_rel_strict} excludes the measure-zero endpoint to prevent spurious equality
cases when the target is reached only at $s=T$.
\end{rem}

\textbf{One-Step Contraction}

We introduce the one step Bellman operator primarily to obtain an operator theoretic fixed point view of the \ac{dpp}; for $\lambda>0$ it yields uniqueness and geometric convergence of value iteration, and the same contraction will be reused in \secref{sec:RL_forward_equivalence} under the forward (time-to-go) parametrization.

Define the backward-time slab and sup norm
\begin{multline}
\mathsf D_\sigma:=\{(t,x)\in[0,T]\times\mathbb R^n:\ t\le T-\sigma\},
\\
\|\Phi\|_\infty:=\sup_{(t,x)\in\mathsf D_\sigma}|\Phi(t,x)|.
\label{eq:domain_norm}
\end{multline}

Let us define a Bellman step \\

\begin{definition}[Bellman step]\label{def:Bellman_step}
For bounded $\Phi:\mathsf D_\sigma\to\mathbb R$ set
\begin{align}
(\mathcal S_{\sigma,\lambda}\Phi)(t,x)
:= \inf_{u\in\mathcal M(t)}\Big\{
&\int_{t}^{t+\sigma}\! e^{\lambda(t-s)}\,
  h\!\big(s,\,x^u_{t,x}(s),\,u(s)\big)\,ds
\notag\\[-2pt]
&\qquad+\,e^{-\lambda\sigma}\,
  \Phi\!\big(t+\sigma,\,x^u_{t,x}(t+\sigma)\big)
\Big\}.
\label{eq:Bellman_step_def}
\end{align}
\end{definition}

\begin{theorem}[Contraction of the Bellman step]\label{thm:contraction_fnl}
For any bounded $\Phi_1,\Phi_2:\mathsf D_\sigma\to\mathbb R$,
\begin{equation}
\|\mathcal S_{\sigma,\lambda}\Phi_1
     -\mathcal S_{\sigma,\lambda}\Phi_2\|_\infty
\ \le\ e^{-\lambda\sigma}\,
      \|\Phi_1-\Phi_2\|_\infty.
\label{eq:contraction_fnl}
\end{equation}
In particular, if $\lambda>0$ then $\mathcal S_{\sigma,\lambda}$ is a strict contraction with modulus $e^{-\lambda\sigma}<1$; if $\lambda=0$ it is nonexpansive.
\end{theorem}

\begin{pf}
Fix $(t,x)\in\mathsf D_\sigma$ and define, for $u\in\mathcal M(t)$,
\begin{multline*}
    F_i(u):=\int_{t}^{t+\sigma} e^{\lambda(t-s)}h(\cdot)\,ds
        + e^{-\lambda\sigma}\Phi_i\big(t+\sigma,X_u\big),
\\ i\in\{1,2\}.
\end{multline*}
Then $(\mathcal S_{\sigma,\lambda}\Phi_i)(t,x)=\inf_{u\in\mathcal M(t)}F_i(u)$.
Using $\inf F_1-\inf F_2 \le \sup_{u}(F_1(u)-F_2(u))$ yields
\begin{align*}
(\mathcal S_{\sigma,\lambda}\Phi_1-\mathcal S_{\sigma,\lambda}\Phi_2)(t,x)
&\le \sup_{u\in\mathcal M(t)} e^{-\lambda\sigma}\,
      \big(\Phi_1-\Phi_2\big)\big(t+\sigma,X_u\big)\\
&\le e^{-\lambda\sigma}\,\|\Phi_1-\Phi_2\|_\infty.
\end{align*}
Exchanging $(\Phi_1,\Phi_2)$ gives the same bound for the negative part, hence
\[
\big|(\mathcal S_{\sigma,\lambda}\Phi_1-\mathcal S_{\sigma,\lambda}\Phi_2)(t,x)\big|
\le e^{-\lambda\sigma}\,\|\Phi_1-\Phi_2\|_\infty.
\]
Taking $\sup_{(t,x)\in\mathsf D_\sigma}$ proves \eqref{eq:contraction_fnl}.
\end{pf}

\begin{rem}[Fixed point]\label{rem:fixed_point}
By \eqref{eq:DPP_rel_fnl_pap}, $V_\lambda$ satisfies $V_\lambda=\mathcal S_{\sigma,\lambda}V_\lambda$ on $\mathsf D_\sigma$.
If $\lambda>0$ and $\sigma>0$, then $\mathcal S_{\sigma,\lambda}$ is a strict contraction on
$(\mathcal B(\mathsf D_\sigma),\|\cdot\|_\infty)$ with modulus $e^{-\lambda\sigma}$.
Hence $V_\lambda$ is the unique fixed point, and for any bounded $\Phi_0$ the iterates
$\Phi_{k+1}:=\mathcal S_{\sigma,\lambda}\Phi_k$ satisfy
\[
\|\Phi_k - V_\lambda\|_\infty \le e^{-\lambda\sigma k}\,\|\Phi_0 - V_\lambda\|_\infty.
\]
\end{rem}

\textit{Note: The Backward formulation can be converted to initial time formulations using same arguments as provided in \cite{evans1984differential}}\\ 

\textbf{Forward (initial-value) formulation.}
\begin{align}
W_\lambda(\tau,x)
&= \inf_{\bar u}\int_{0}^{\tau}\! e^{-\lambda r}\,
     h\!\big(T-\tau+r,\,y(r),\,\bar u(r)\big)\,dr,
\label{eq:W_val}
\\[-1pt]
W_\lambda(0,x)&=0.
\notag
\end{align}

The \ac{dpp} (for $\sigma\in[0,\tau]$) reads
\begin{align}
W_\lambda(\tau,x)
&= \inf_{\bar u}\Big\{
\int_{0}^{\sigma}\! e^{-\lambda r}\,
  h\!\big(T-\tau+r,\,y(r),\,\bar u(r)\big)\,dr
\notag\\[-1pt]
&\hspace{6.6em}
+\,e^{-\lambda\sigma}\,
   W_\lambda\!\big(\tau-\sigma,\,y(\sigma)\big)
\Big\}.
\label{eq:DPP_forward}
\end{align}
Hamilton-Jacobi-Bellman (initial value problem):
\begin{multline}
W_{\lambda,\tau}(\tau,x)
- H\!\big(T-\tau,x,\nabla_x W_\lambda(\tau,x)\big)
+ \lambda\,W_\lambda(\tau,x)=0,
\\ W_\lambda(0,x)=0.
\label{eq:HJB_forward}
\end{multline}

\section{HJB reachability and RL Equivalence}
\label{sec:RL_forward_equivalence}

We now rewrite the discounted dynamic-programming principle in forward time and interpret it as the Bellman equation of a deterministic discounted MDP. This is not a new Bellman construction: it is the time-reversed form of the discounted Bellman step introduced in Section~\ref{sec:discounted_hjb}, now written in state-action-next-state form to match the standard RL viewpoint.

Fix a step size $\sigma\in(0,T]$ and $\lambda\ge0$. For each $(\tau,x)\in[\sigma,T]\times\R^n$:

\smallskip
\noindent\textbf{State.} $(\tau,x)$.

\noindent\textbf{Action on one step.}
Any measurable control segment
$a:[0,\sigma]\to\mathcal U$. Denote the set of such segments by
$\mathcal A_\sigma$.

\noindent\textbf{Step dynamics.}
Let $y(\cdot)$ solve
\begin{equation}\label{eq:step_dyn}
y'(r)=f\big(y(r),a(r)\big),\qquad y(0)=x,\qquad r\in[0,\sigma],
\end{equation}
and set the next state to $(\tau-\sigma,y(\sigma))$.

\noindent\textbf{Per-step discounted cost.}
\begin{equation}\label{eq:step_cost}
c(\tau,x,a)
:= \int_{0}^{\sigma}\! e^{-\lambda r}\,
   h\!\big(T-\tau+r,\,y(r),\,a(r)\big)\,dr.
\end{equation}

\noindent\textbf{Discount factor.}
\[
\gamma:=e^{-\lambda\sigma}\in(0,1].
\]

The corresponding forward Bellman operator on bounded
$\Psi:[0,T]\times\R^n\to\R$ is
\begin{align}
(\mathcal T_{\sigma,\lambda}\Psi)(\tau,x)
:= \inf_{a\in\mathcal A_\sigma}\Big\{&
c(\tau,x,a)\ +\ \gamma\,
\Psi\big(\tau-\sigma,\,y(\sigma)\big)\Big\}.
\label{eq:Bellman_forward_def}
\end{align}

Because \eqref{eq:Bellman_forward_def} uses the exact ODE flow over $[0,\sigma]$ and the exact discounted running cost on that interval, it is an exact Bellman reformulation of the continuous time problem. In particular, the forward DPP \eqref{eq:DPP_forward} implies
\begin{equation}
W_\lambda(\tau,x)=(\mathcal T_{\sigma,\lambda}W_\lambda)(\tau,x),
\qquad \forall(\tau,x)\in[\sigma,T]\times\R^n.
\label{eq:forward_fixed_point_exact}
\end{equation}

\begin{theorem}[Contraction and fixed point uniqueness]
\label{thm:forward_contraction_uniqueness}
For bounded $\Psi_1,\Psi_2:[0,T]\times\R^n\to\R$,
\begin{equation}\label{eq:forward_contraction}
\|\mathcal T_{\sigma,\lambda}\Psi_1-\mathcal T_{\sigma,\lambda}\Psi_2\|_\infty
\ \le\ e^{-\lambda\sigma}\,\|\Psi_1-\Psi_2\|_\infty.
\end{equation}
Hence, if $\lambda>0$, $\mathcal T_{\sigma,\lambda}$ is a strict contraction on bounded functions over $[\sigma,T]\times\R^n$ and therefore has a unique fixed point. Moreover,
\begin{equation}\label{eq:RL_equals_W}
W_\lambda=\mathcal T_{\sigma,\lambda}W_\lambda,
\qquad
\lim_{k\to\infty}\mathcal T_{\sigma,\lambda}^k \Psi
= W_\lambda
\end{equation}
for every bounded initial seed $\Psi$, with geometric rate $e^{-\lambda\sigma}$.
\end{theorem}

\begin{pf}
The proof is identical in structure to Theorem~\ref{thm:contraction_fnl}. The step cost \(c(\tau,x,a)\) cancels when comparing \((\mathcal T_{\sigma,\lambda}\Psi_1)(\tau,x)\) and \((\mathcal T_{\sigma,\lambda}\Psi_2)(\tau,x)\), leaving only the discounted continuation term. Thus, for any fixed \((\tau,x)\),
\[
\big|(\mathcal T_{\sigma,\lambda}\Psi_1-\mathcal T_{\sigma,\lambda}\Psi_2)(\tau,x)\big|
\le e^{-\lambda\sigma}\|\Psi_1-\Psi_2\|_\infty.
\]
Taking the supremum over \((\tau,x)\) gives \eqref{eq:forward_contraction}. If $\lambda>0$, then $e^{-\lambda\sigma}<1$, so Banach's fixed-point theorem yields uniqueness and geometric convergence. The identity \(W_\lambda=\mathcal T_{\sigma,\lambda}W_\lambda\) follows directly from \eqref{eq:forward_fixed_point_exact}.
\end{pf}

\begin{rem}[RL interpretation]
The operator \eqref{eq:Bellman_forward_def} is the Bellman operator of a deterministic discounted MDP with state \((\tau,x)\), intra-step action \(a(\cdot)\), per-step cost \eqref{eq:step_cost}, and discount factor \(e^{-\lambda\sigma}\). Thus, for \(\lambda>0\), standard value or policy iteration converges to \(W_\lambda\) for this exact one-step MDP.
\end{rem}

\subsection{PDE limit for implementable one-step schemes}
\label{subsec:PDE_limit_numerical}

In the text above, we constructed an exact $\sigma$ step Bellman operator by using the exact ODE flow
and the exact discounted running cost over $[0,\sigma]$. Consequently we proved that the $W_\lambda$ is its fixed point for every $\sigma$. In practice, RL implementations use a numerical one-step model. The state transition is computed
by a time-stepping integrator (e.g.\ Euler/RK) and the step cost is computed by a quadrature rule \cite{golub1969calculation}. We now show that the resulting discrete Bellman fixed points converge to the viscosity solution of the forward HJB as $\sigma\downarrow 0$.

\textbf{Numerical one-step model}\\
Fix $\sigma\in(0,T]$ and $\lambda\ge 0$. On each step we restrict actions to be constant controls
$u\in\mathcal U$ (piecewise-constant policies across steps), which matches standard discrete time RL.

Let $\widehat F_\sigma:\R^n\times\mathcal U\to\R^n$ be a one-step numerical integrator for $\dot y=f(y,u)$.
For example, explicit Euler gives $\widehat F_\sigma(x,u)=x+\sigma f(x,u)$, and RK schemes give higher-order maps.
Let $\widehat c_{\sigma,\lambda}:[0,T]\times\R^n\times\mathcal U\to\R$ be a one-step cost approximation
(e.g.\ a Riemann or quadrature approximation of $\int_0^\sigma e^{-\lambda r}h(T-\tau+r,y(r),u)\,dr$).

We assume the following local consistency holds uniformly on compact subsets:
\begin{align}
\widehat F_\sigma(x,u) &= x+\sigma f(x,u)+o(\sigma),
\label{eq:num_flow_consistency}
\\
\widehat c_{\sigma,\lambda}(\tau,x,u) &= \sigma\,h(T-\tau,x,u)+o(\sigma),
\label{eq:num_cost_consistency}
\end{align}
as $\sigma\downarrow 0$, uniformly for $(\tau,x,u)$ in compact sets. Moreover, we assume $\widehat c_{\sigma,\lambda}$ is bounded whenever $h$ is bounded.

The results below apply to any one-step integrator/quadrature pair
$(\widehat F_\sigma,\widehat c_{\sigma,\lambda})$ satisfying the
consistency conditions \eqref{eq:num_flow_consistency}-\eqref{eq:num_cost_consistency}
(and boundedness). For example explicit Euler with a left-Riemann (or trapezoidal) cost
approximation.

\begin{definition}[Numerical Bellman operator]
\label{def:num_Bellman}
For bounded $\Psi:[0,T]\times\R^n\to\R$, define
\begin{multline}
(\widehat{\mathcal T}_{\sigma,\lambda}\Psi)(\tau,x)
:= \\ \inf_{u\in\mathcal U}\Big\{
\widehat c_{\sigma,\lambda}(\tau,x,u)\ +\ e^{-\lambda\sigma}\,
\Psi\big(\tau-\sigma,\widehat F_\sigma(x,u)\big)\Big\}
\label{eq:num_Bellman_def}
\end{multline}
for $(\tau,x)\in[\sigma,T]\times\R^n$, with boundary data $\Psi(\tau,x)=0$ on $\tau\in[0,\sigma)$.
Let $W^\sigma$ denote the fixed point of $\widehat{\mathcal T}_{\sigma,\lambda}$.
\end{definition}

\begin{rem}[Existence/uniqueness when $\lambda>0$]
\label{rem:num_contraction}
The proof of Theorem~\ref{thm:forward_contraction_uniqueness} applies verbatim to
$\widehat{\mathcal T}_{\sigma,\lambda}$ since the dependence on $\Psi$ is still only through the term
$e^{-\lambda\sigma}\Psi(\cdot)$. Hence
\[
\|\widehat{\mathcal T}_{\sigma,\lambda}\Psi_1-\widehat{\mathcal T}_{\sigma,\lambda}\Psi_2\|_\infty
\le e^{-\lambda\sigma}\|\Psi_1-\Psi_2\|_\infty.
\]
If $\lambda>0$, $\widehat{\mathcal T}_{\sigma,\lambda}$ is a strict contraction and the fixed point $W^\sigma$ is unique.
\end{rem}

Now we will prove Theorem \ref{thm:num_BS_forward} using the the Barles-Souganidis \cite{barles1991convergence} framework. This is the rigorous bridge proving that as $\sigma\to 0$, the discrete RL fixed points $W^\sigma$ converge to the continuous time value $W_\lambda$.

\begin{theorem}[Convergence to the viscosity solution]
\label{thm:num_BS_forward}
Assume $\lambda>0$ and the standing regularity, and let $W^\sigma$ be the unique fixed point of
$\widehat{\mathcal T}_{\sigma,\lambda}$ (Definition~\ref{def:num_Bellman}).
Then, as $\sigma\downarrow 0$,
\[
W^\sigma \ \to\ W_\lambda \qquad \text{locally uniformly on } [0,T]\times\R^n,
\]
where $W_\lambda$ is the unique bounded viscosity solution of the forward HJB.
\end{theorem}
\begin{proofver}
\begin{pf}
By Lemmas shown in Appendix~\ref{lem:num_mono_stable_appen}-\ref{lem:num_consistency_appen}, the numerical scheme is monotone,
stable, and consistent with the forward HJB. Since the forward HJB is proper for $\lambda>0$,
comparison holds for bounded viscosity solutions, and the Barles Souganidis theorem
\cite{barles1991convergence} yields local uniform convergence of $W^\sigma$ to the unique viscosity solution,
which is $W_\lambda$.
\end{pf}
\end{proofver}
\begin{ideaver}
\begin{pf}
    The numerical scheme is can be shown to be monotone, stable, and consistent with the forward HJB (for further detail see the arXiv version of the paper \cite{solanki2026formalizingrelationshiphamiltonjacobireachability}). Since the forward HJB is proper for $\lambda>0$, comparison holds for bounded viscosity solutions, and the Barles Souganidis theorem \cite{barles1991convergence} yields local uniform convergence of $W^\sigma$ to the unique viscosity solution, which is $W_\lambda$.
\end{pf}
\end{ideaver}

Thus we have that the discrete RL Bellman update \eqnref{eq:Bellman_forward_def} is a
provably consistent, monotone, stable approximation of the forward \ac{hjb}.
Value iteration converges (for $\lambda>0$) to $W_\lambda$, and as the
step $\sigma\to0$ the discrete fixed points $W^\sigma$ converge to the
viscosity solution of the PDE.

\begin{proofver}
Now we will show that the Bellman residual used in RL training matches, in the small step limit, the PDE residual. it justifies using Bellman-residual minimization as a proxy for solving the \ac{hjb} and explains why driving the residual to zero enforces the correct continuous time optimality conditions.

For $\phi\in C^1$, define the numerical Bellman residual
\[
\widehat{\mathcal R}_{\sigma,\lambda}[\phi](\tau,x)
:= \frac{\phi(\tau,x)-(\widehat{\mathcal T}_{\sigma,\lambda}\phi)(\tau,x)}{\sigma}.
\]
Then Lemma in Appendix~\ref{lem:num_consistency_appen} immediately implies
\[
\widehat{\mathcal R}_{\sigma,\lambda}[\phi](\tau,x)
\ \xrightarrow[\ \sigma\downarrow 0\ ]{}\ 
\phi_\tau(\tau,x)-\widetilde H(\tau,x,\nabla\phi(\tau,x))
+\lambda\,\phi(\tau,x),
\]
uniformly on compact subsets. Thus minimizing the Bellman residual in the small-step regime
targets the HJB residual.
\end{proofver}
\begin{ideaver}
    It can be shown that the Bellman residual used in RL training matches, in the small step limit, the PDE residual. it justifies using Bellman-residual minimization as a proxy for solving the \ac{hjb} and explains why driving the residual to zero enforces the correct continuous time optimality conditions. For proof please refer to the arXiv version of the paper \cite{solanki2026formalizingrelationshiphamiltonjacobireachability}.
    \[
    \widehat{\mathcal R}_{\sigma,\lambda}[\phi](\tau,x)
    \ \xrightarrow[\ \sigma\downarrow 0\ ]{}\ 
    \phi_\tau(\tau,x)-\widetilde H(\tau,x,\nabla\phi(\tau,x))
    +\lambda\,\phi(\tau,x),
    \]
\end{ideaver}

\begin{rem}[Intuition]
At smooth test functions, the RL Bellman residual equals (in the small
step limit) the \ac{hjb} residual. Hence the PDE encodes the fixed point
condition of the Bellman operator in continuous time.
\end{rem}

\section{Methodology and Experiments}
\label{sec:method}

We validate the proposed bridge between Hamilton-Jacobi (\ac{hj}) reachability and reinforcement learning (RL) in two stages. Throughout, the physical system is the double integrator
\begin{equation}
\dot x_1 = x_2,\qquad \dot x_2 = u,\qquad u\in\{-a_{\max},+a_{\max}\},
\label{eq:double-int}
\end{equation}
with $a_{\max}>0$. The target set is the open disk of radius $r$,
\begin{equation}
\mathcal T:=\{x \in \mathbb{R}^{2}:\,\|x\|<r\},
\label{eq:target-set}
\end{equation}
and the travel cost encodes target membership via
\begin{equation}
  h(x,u) \;=\;
  \begin{cases}
    -\alpha (r-\|x\|), & \|x\|< r,\\[2pt]
    0,       & \|x\|\ge r,
  \end{cases}
  \qquad \alpha>0,\; r>0,
  \label{eq:travel-cost}
\end{equation}
where $\alpha$ is a scaling factor. This sign calibration ($h\equiv 0$ off target and $h<0$ on target) is crucial for recovering strict reachability from level sets of the value.

\begin{proofver}
\textbf{Relation to the theoretical setup.}
The double-integrator dynamics in \eqnref{eq:double-int} are the physical dynamics used in the experiments. As discussed in Section~\ref{sec:problem_statement}, the Hamilton Jacobi analysis is stated for a globally defined auxiliary vector field that agrees with the physical dynamics on an open neighborhood of the reported regions of interest (ROIs). Since all numerical evaluations reported in this section are carried out on those ROIs, the theoretical and physical dynamics coincide throughout the domain relevant to the experiments.
\end{proofver}

\subsection{Stage I: Travel-vs-Reach \ac{hjb} (zero/negative level set equivalence)}
\label{subsec:method-stage1}

We compare two \ac{hjb} formulations on a common grid over a fixed ROI:
\begin{enumerate}[leftmargin=2.2em,label=(\roman*)]
\item \textbf{Classical reach cost (minimum-over-time)}, leading to the standard \ac{hj} reachability PDE and strict backward-reachable tube (\ac{brt}).
\item \textbf{Travel cost (\eqnref{eq:travel-cost})}, leading to an \ac{hjb} value whose negative sublevel set equals the strict \ac{brt} and whose zero level set coincides with the \ac{brt} boundary.
\end{enumerate}
For this experiment, we used the existing reachability toolboxes HelperOC and the Level Set Methods Toolbox~\cite{choi2021robust,mitchell2008flexible}.

\subsection{Stage II: Forward discounted \ac{hjb} $\leftrightarrow$ RL with continuation}
\label{subsec:method-stage2}

We relate a discounted forward \ac{hjb} to an RL fixed point via a monotone, stable, and consistent time discretization.

\textbf{Discounted stationary \ac{hjb}.}
For a discount rate $\lambda>0$, we consider the stationary discounted \ac{hjb} associated with the double-integrator dynamics and compute its numerical approximation on a bounded Cartesian grid over the ROI:
\begin{align}
\lambda\,V(x) &= \min_{u\in\{u_L,u_H\}}\Big\{\,h(x,u)+\nabla V(x)\!\cdot\! f(x,u)\,\Big\},
\label{eq:hjb-stationary}\\
f(x,u) &= \begin{pmatrix}x_2\\u\end{pmatrix},\qquad
u_L=-a_{\max},\qquad
u_H=+a_{\max}.
\nonumber
\end{align}
We solve this PDE baseline via a first-order semi-Lagrangian fixed-point iteration on a uniform grid. Over a short step $\Delta\tau$, the discounted Bellman map is discretized as
\begin{equation}
\begin{aligned}
(\mathcal{T}V)(x)
&= \min_{u\in\{u_L,u_H\}}
\Big\{\, w\,h(x,u)
+ \gamma\,V\!\big(x+\Delta\tau f(x,u)\big)\Big\}, \\
\gamma &= e^{-\lambda\Delta\tau}, \qquad
w = \tfrac{1-\gamma}{\lambda}.
\end{aligned}
\label{eq:bellman-discrete-stationary}
\end{equation}
We use a forward Euler step for the characteristic and a first-order approximation of the running-cost integral through the weight $w=(1-\gamma)/\lambda$. For the PDE baseline only, off-grid state queries arising in the semi-Lagrangian continuation term are clamped back to the computational boundary. This is a numerical boundary treatment for the bounded-domain comparison solver and is distinct from the theoretical formulation in Sections~\ref{sec:problem_statement}-\ref{sec:discounted_hjb}. We perform synchronous value iteration $V^{k+1}=\mathcal{T}V^{k}$ until the sup-norm change falls below $10^{-6}$ or a cap of $2000$ iterations is reached. The resulting discretization is monotone, stable, and consistent; accordingly, by the Barles-Souganidis framework \cite{barles1991convergence}, its fixed points converge to the viscosity solution of \eqnref{eq:hjb-stationary} as the temporal and spatial discretization steps tend to zero~\cite{bardi1997optimal,falcone2013semi,barles1991convergence,crandall1992user}.

\textbf{RL training (fitted value).}
We train a value network $W_\theta(x)$ to approximate the forward discounted value function using a \ac{td} loss. The input is the state $(x_1,x_2)$, and the network outputs a single scalar $W_\theta(x)$ estimating the discounted cumulative cost-to-go. The TD target includes a minimization over the bang bang control actions and discount factor $\gamma=e^{-\lambda\Delta\tau}$. We use a two-layer \ac{siren} with 256 neurons per hidden layer and base frequency $30~\mathrm{rad/s}$~\cite{sitzmann2020implicit}, following the representation choice used in DeepReach~\cite{bansal2021deepreach}. Training is sample-based: one-step transitions are generated by simulation under the discrete controls \(u_L,u_H\), and the network is updated from the resulting state transition cost tuples. The full training loop is provided in Appendix~\ref{appen:rl_training_loop}.

\subsection{Evaluation protocol (common to both stages)}
\label{subsec:method-eval}

All comparisons are conducted on uniform Cartesian grids over task-specific ROIs:
\begin{itemize}[leftmargin=2.0em]
\item ROI for Stage~I (travel vs.\ reach): $\mathcal X_{10}=[-10,10]\times[-10,10]$.
\item ROI for Stage~II (\ac{hjb} $\leftrightarrow$ RL): $\mathcal X_{2.5}=[-2.5,2.5]\times[-2.5,2.5]$.
\end{itemize}
We use uniform Cartesian grids of sizes $501\times 501$ and $201\times 201$ on the two ROIs, respectively. For Stage~I, we compare zero and negative level sets of the reach-cost and travel-cost formulations. For Stage~II, we report the maximum and mean absolute errors between the neural value and the PDE solution on the same evaluation grid. In the discounted experiments, we use $\Delta\tau=0.05$ and $\lambda=1.0$, and these values are kept identical between the PDE and RL targets in Stage~II.
\section{Results}
\label{sec:results}

\begin{figure*}[t]
\centering
\subfloat[Reach-cost value (zero-level contour shown).]{
    \includegraphics[width=0.31\textwidth]{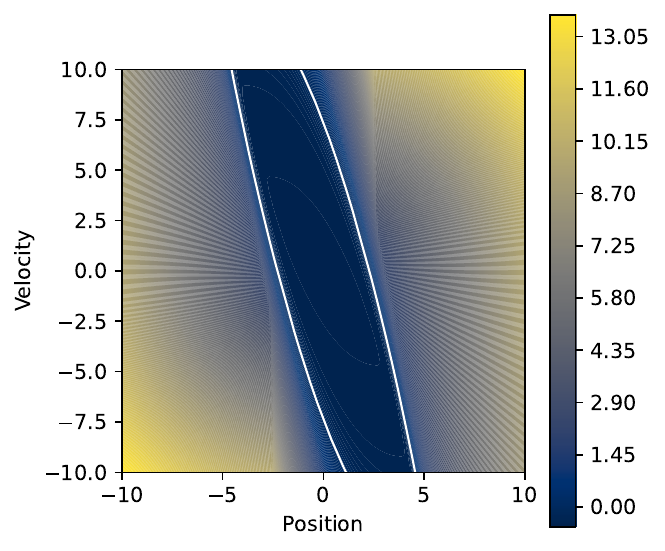}
    \label{fig:reach}
}\hfill
\subfloat[Travel-cost value (same grid and solver).]{
    \includegraphics[width=0.31\textwidth]{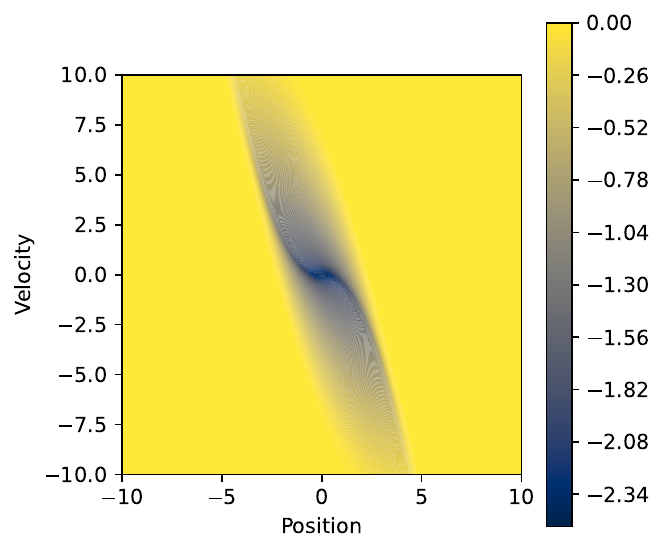}
    \label{fig:travel}
}\hfill
\subfloat[Histogram of travel-cost values inside the reach-cost zero-level set.]{
    \includegraphics[width=0.31\textwidth]{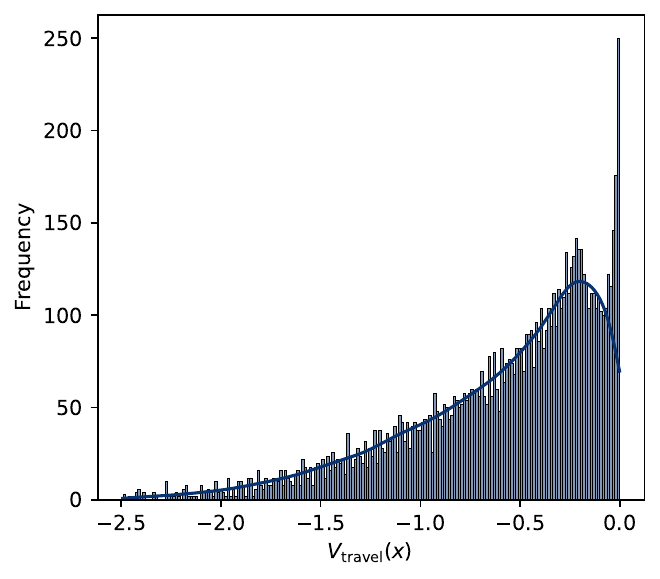}
    \label{fig:hist}
}
\caption{Travel- vs.\ reach-cost \ac{hjb} solutions computed on $\mathcal{X}_{10}$ for double integrator.} 
\label{fig:reach-travel-di}
\end{figure*}

\begin{figure*}[t]
\centering
\subfloat[Discounted \ac{hjb} (PDE) solution.]{
    \includegraphics[width=0.32\textwidth]{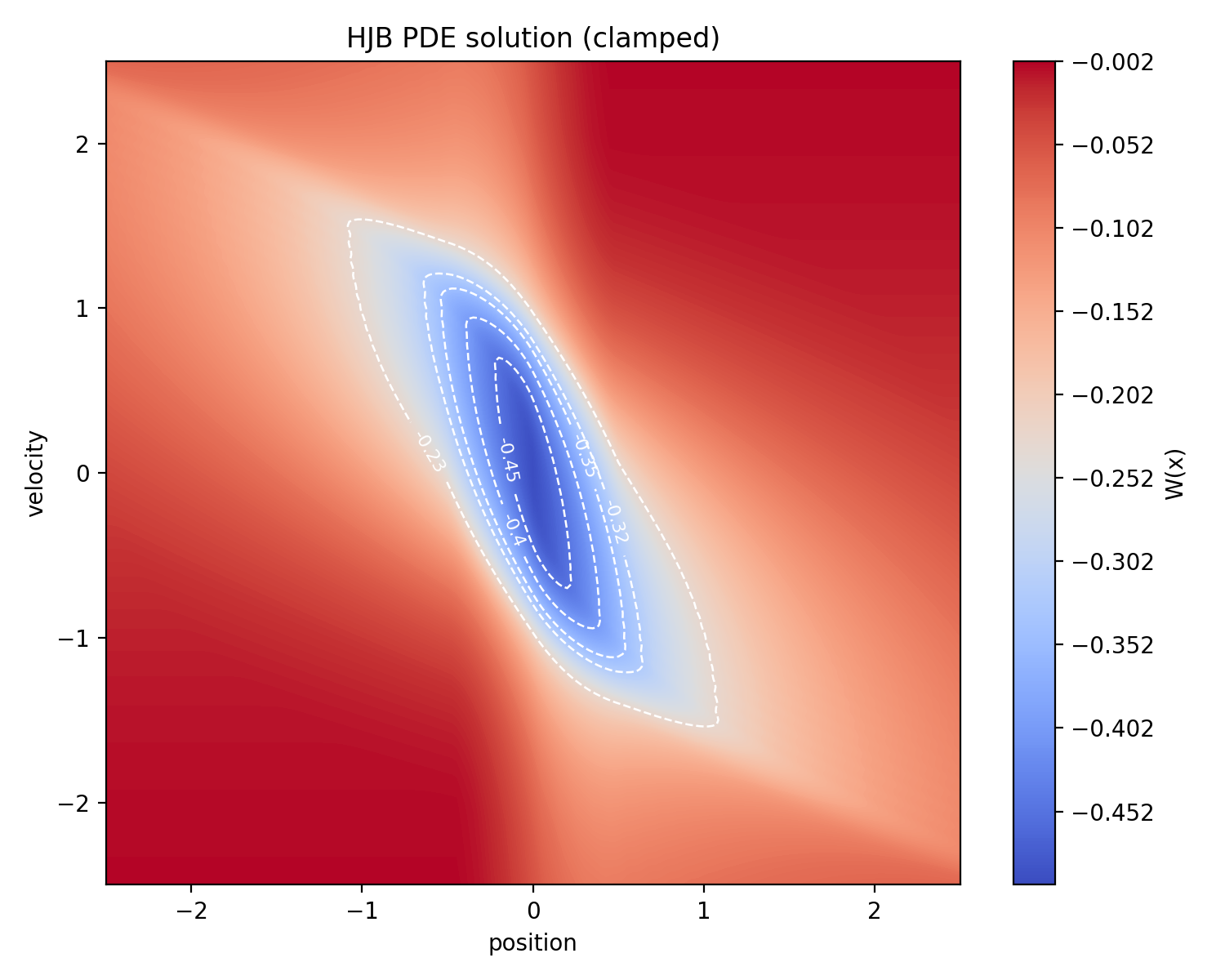}
    \label{fig:nn-pde-pde}
}\hfill
\subfloat[Learned value (NN).]{
    \includegraphics[width=0.32\textwidth]{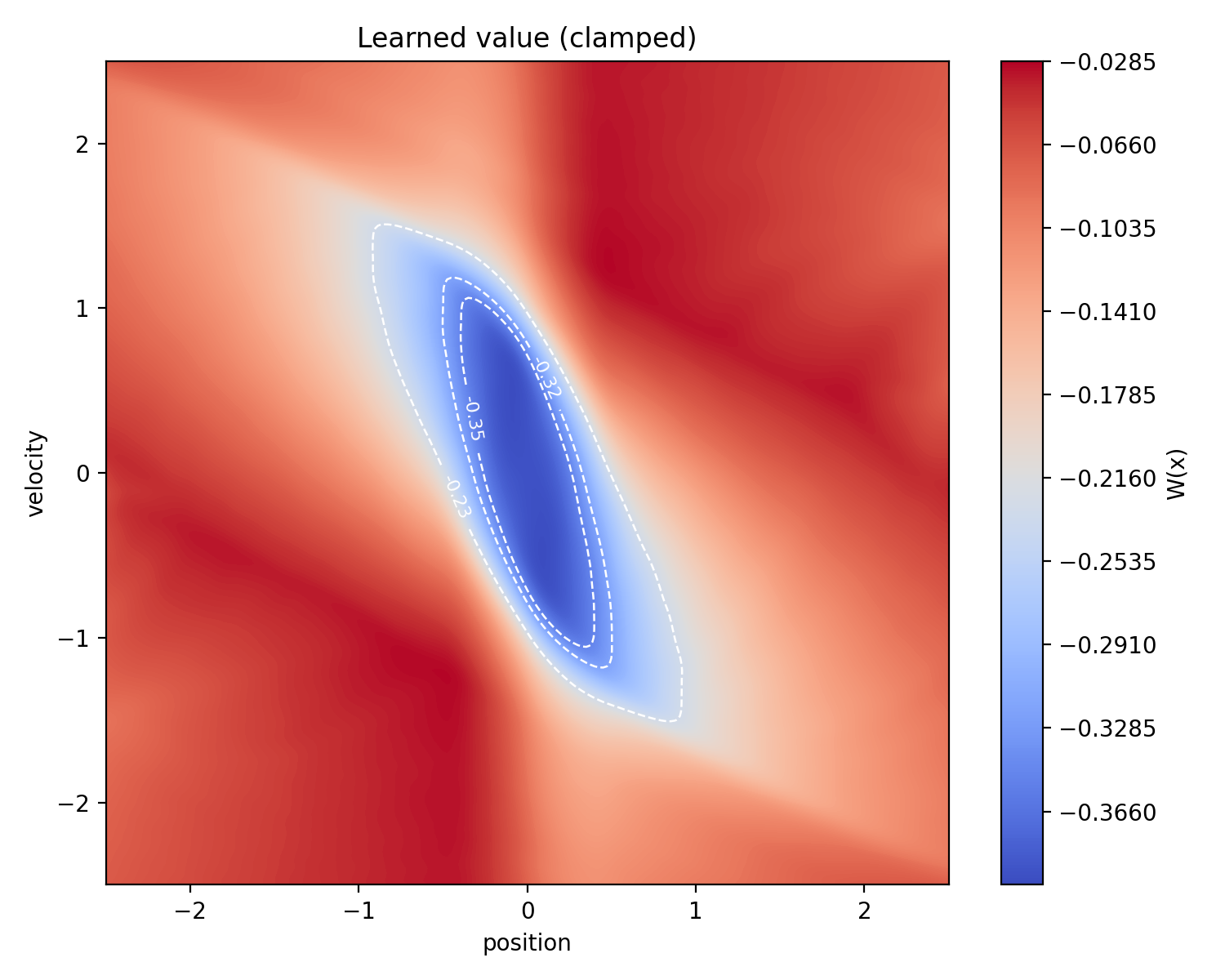}
    \label{fig:nn-pde-learn}
}\hfill
\subfloat[Error field $V_{\text{PDE}}-W_\theta$.]{
    \includegraphics[width=0.32\textwidth]{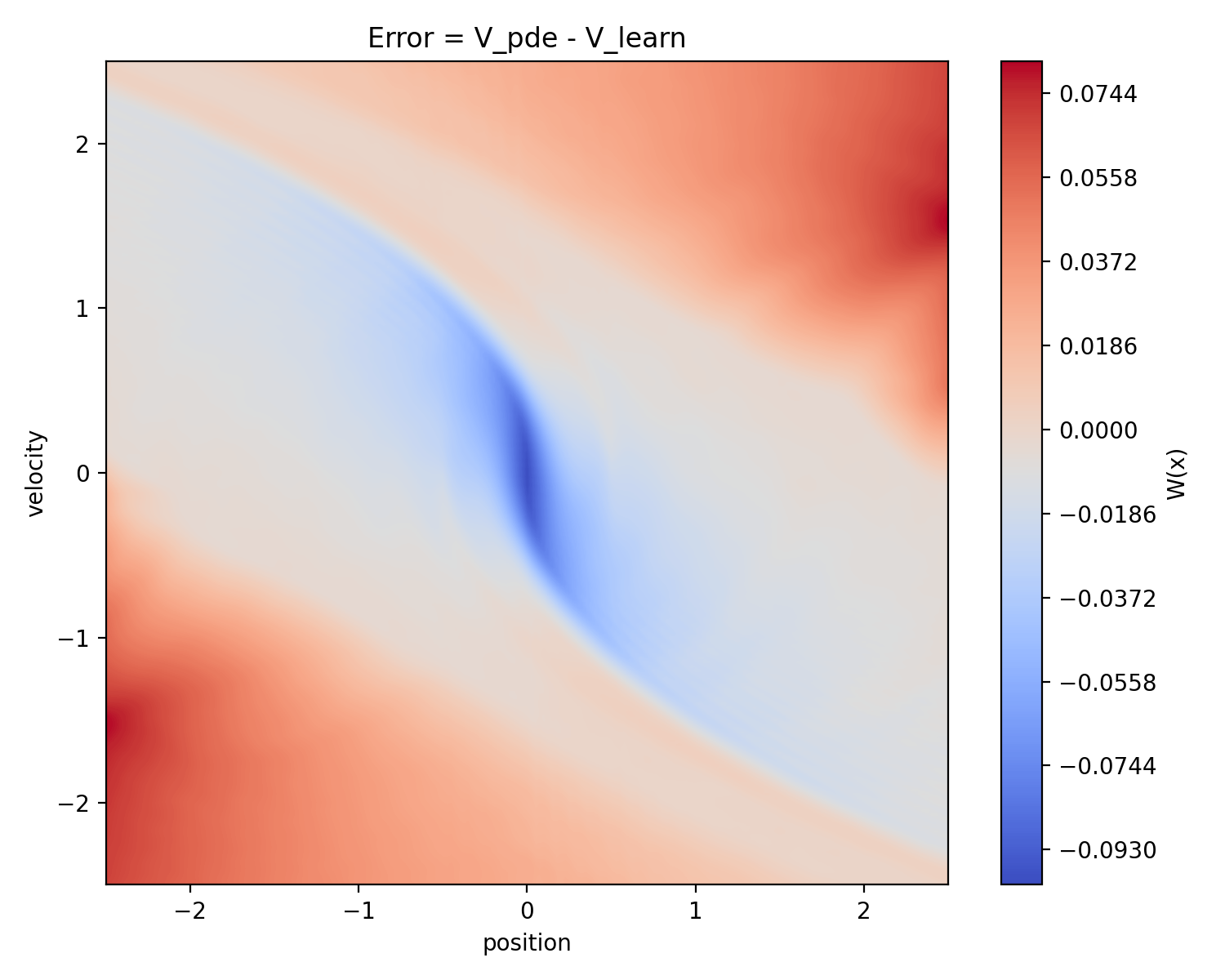}
    \label{fig:nn-pde-error}
}

\caption{Forward discounted \ac{hjb} $\leftrightarrow$ RL on 
$\mathcal{X}_{2.5}=[-2.5,2.5]^2$ with $\Delta\tau=0.05$, 
$\lambda=1.0$ ($\gamma=e^{-0.05}$). Visual agreement is strong across the ROI; 
quantitative errors are reported in \eqnref{eq:errors}.}
\label{fig:nn-pde-di}
\end{figure*}

\subsection{Stage I: Travel-cost \ac{hjb} reproduces strict \ac{brt}}
On $\mathcal{X}_{10}$, the travel-cost \ac{hjb} defined by \eqnref{eq:travel-cost} yields a value function whose negative sublevel set coincides with the strict backward-reachable tube (\ac{brt}), while its complement corresponds to the zero level set; see \figref{fig:reach-travel-di}. This shows that strict reachability can be encoded by a purely running-cost formulation, without a terminal penalty. Because the travel-cost value saturates at zero outside the reachable region, the zero level set is numerically degenerate and cannot be extracted directly. To visualize the correspondence, we overlay the reach-cost zero-level contour on the travel-cost field and inspect the interior values (\figref{fig:hist}), which are all strictly negative.

\subsection{Stage II: Forward discounted \ac{hjb} matches RL with continuation}
On $\mathcal X_{2.5}$, we compare the learned value $W_\theta$ with the discounted semi-Lagrangian \ac{hjb} solution $V$ on the same grid. For $\Delta\tau=0.05$ and $\lambda=1.0$ (so $\gamma=e^{-0.05}$), the agreement is
\begin{equation}
\label{eq:errors}
  \max\nolimits_{\text{grid}} \bigl| W_\theta - V \bigr| \approx 0.1006,\qquad
  \mathbb{E}_{\text{grid}} \bigl| W_\theta - V \bigr| \approx 0.0215.
\end{equation}
Representative heatmaps of the PDE solution (\figref{fig:nn-pde-pde}) and the learned neural-network value (\figref{fig:nn-pde-learn}) are shown, with the corresponding error field in \figref{fig:nn-pde-error}.

Additional results are reported in the appendix \ref{app:higher_dimensional_results} and \ref{app:lambda_sensitivity}. First, a \(\lambda\)-sensitivity sweep on the 2D double-integrator benchmark shows that larger \(\lambda\) leads to faster TD-loss decay, consistent with the stronger contraction factor \(e^{-\lambda\Delta\tau}\), while also changing the underlying discounted objective. Second, we include higher-dimensional experiments on a 4D planar double integrator and a 6D integrator. In both cases, training is performed using TD loss only, while an HJB loss is evaluated numerically during training. The observed decrease of both losses provides additional evidence that the TD-trained value moves toward the HJB solution beyond the 2D benchmark.
\section{Conclusion and Future Work}
\label{sec:conclusion}

We established a bridge between Hamilton--Jacobi (\ac{hj}) reachability and reinforcement learning (RL) by expressing reachability in an RL-compatible dynamic-programming form. A calibrated travel-cost \ac{hjb} exactly recovers strict reachability, a discounted forward formulation yields an exact Bellman fixed-point equation and a contraction for \(\lambda>0\), and a semi-Lagrangian discretization provides a monotone, stable, and consistent scheme converging to the viscosity solution in the small-step limit.

Looking ahead, we aim to extend the framework to reach--avoid games with Isaacs operators and disturbances, stochastic and risk-sensitive settings, and safe on-policy exploration.


\bibliographystyle{plain}        
\bibliography{articles}    

\appendix

\section{DPP with relative discount}\label{lem:DPP_rel_fnl_appen}
\begin{proofver}
\begin{lemmastar}[\ac{dpp} with relative discount]
For any $(t,x)\in[0,T]\times\mathbb R^n$ and $\sigma\in[0,\,T-t]$,
\begin{align}
V_\lambda(t,x)
&= \inf_{u\in\mathcal M(t)}\Bigg\{
\int_{t}^{t+\sigma}\! e^{\lambda(t-s)}\,
   h\!\big(s,\,x^u_{t,x}(s),\,u(s)\big)\,ds
\notag\\[-1pt]&\hspace{7.45em}
+\,e^{-\lambda\sigma}\,
   V_\lambda\!\big(t+\sigma,\,x^u_{t,x}(t+\sigma)\big)
\Bigg\}.
\label{eq:DPP_rel_fnl}
\end{align}
\end{lemmastar}
\begin{pf}

\textit{Preliminaries.}
Based on \assref{a:f_lipschitz} (and measurability of $u$), the trajectory $x^u_{t,x}$ is unique and continuous. Based on Assumption~\ref{a:h_uniform} and the assumed uniform continuity of $s\mapsto h(s,x,u)$,
the map $s\mapsto h\big(s,x^u_{t,x}(s),u(s)\big)$ is measurable and bounded, hence integrable.
\smallskip
\end{proofver}

\begin{ideaver}
    The proof of Lemma \ref{lem:DPP_rel_fnl}
    \begin{pf}
\end{ideaver}
\noindent\textit{($\le$)} Fix $u\in\mathcal M(t)$ and set $y:=x^u_{t,x}(t+\sigma)$. For $\varepsilon>0$ pick
$v_\varepsilon\in\mathcal M(t+\sigma)$ with
\begin{align}
J_\lambda(t+\sigma,y;v_\varepsilon)
\;\le\; V_\lambda(t+\sigma,y)+\varepsilon.
\label{eq:eps_opt_tail}
\end{align}
Let $w:=u\oplus_{t+\sigma}v_\varepsilon\in\mathcal M(t)$. Then $x^w_{t,x}=x^u_{t,x}$ on $[t,t+\sigma]$
and $x^w_{t,x}=x^{v_\varepsilon}_{t+\sigma,y}$ on $[t+\sigma,T]$, hence
\begin{align}
J_\lambda(t,x;w)
&=\int_{t}^{t+\sigma} e^{\lambda(t-s)}h(\cdot)\,ds
 + \int_{t+\sigma}^{T} e^{\lambda(t-s)}h(\cdot)\,ds
\notag\\[-2pt]
&=\int_{t}^{t+\sigma} e^{\lambda(t-s)}h(\cdot)\,ds
 + e^{-\lambda\sigma}\,J_\lambda(t+\sigma,y;v_\varepsilon),
\label{eq:split_tail}
\end{align}
using $e^{\lambda(t-s)}=e^{-\lambda\sigma}e^{\lambda((t+\sigma)-s)}$ for $s\ge t+\sigma$.
By $V_\lambda(t,x)\le J_\lambda(t,x;w)$ and \eqnref{eq:eps_opt_tail}-\eqnref{eq:split_tail},
\begin{align*}
V_\lambda(t,x)
\le \int_{t}^{t+\sigma} e^{\lambda(t-s)}h(\cdot)\,ds
 + e^{-\lambda\sigma}V_\lambda(t+\sigma,y)+e^{-\lambda\sigma}\varepsilon.
\end{align*}
Infimize over $u\in\mathcal M(t)$ and let $\varepsilon\downarrow 0$.

\smallskip
\noindent\textit{($\ge$)}
Fix $\varepsilon>0$ and choose $u_\varepsilon\in\mathcal M(t)$ so that
\begin{align}
J_\lambda(t,x;u_\varepsilon)\le V_\lambda(t,x)+\varepsilon.
\label{eq:eps_opt_head}
\end{align}
Let $y_\varepsilon:=x^{u_\varepsilon}_{t,x}(t+\sigma)$. Then
\begin{align*}
&J_\lambda(t,x;u_\varepsilon)
\\&=\int_{t}^{t+\sigma} e^{\lambda(t-s)}h(\cdot)\,ds
 + e^{-\lambda\sigma}\,J_\lambda\!\big(t+\sigma,y_\varepsilon;
   u_\varepsilon|_{[t+\sigma,T]}\big)\\
&\ge \int_{t}^{t+\sigma} e^{\lambda(t-s)}h(\cdot)\,ds
 + e^{-\lambda\sigma}\,V_\lambda(t+\sigma,y_\varepsilon).
\end{align*}
Combine with \eqnref{eq:eps_opt_head}, take $\inf_{u\in\mathcal M(t)}$ on the RHS, and send $\varepsilon\downarrow0$.
\end{pf}

\begin{proofver}
\section{Boundedness}\label{lem:bounded_appen}
\begin{lemmastar}[Boundedness]
Under \assref{a:h_uniform},
\begin{multline}
|V_\lambda(t,x)|\ \le\ 
\int_{0}^{T-t}\! e^{-\lambda r}M_h\,dr
= \\ \begin{cases}
\frac{M_h}{\lambda}\big(1-e^{-\lambda(T-t)}\big), & \lambda>0,\\[4pt]
M_h\,(T-t), & \lambda=0.
\end{cases}
\label{eq:bound_value}
\end{multline}
\end{lemmastar}
\begin{pf}
Assume $\lambda\ge 0$. Fix $(t,x)$ and any admissible control $u(\cdot)\in\mathcal M(t)$. 
By \assref{a:h_uniform},
\[
\big|h\big(s,x^u_{t,x}(s),u(s)\big)\big|\le M_h
\qquad\text{for a.e. }s\in[t,T].
\]
Hence, using \eqref{eq:Jlambda_def} and the change of variables $r:=s-t$,
\begin{align*}
|J_\lambda(t,x;u)|
&=\left|\int_t^T e^{\lambda(t-s)}\,h\big(s,x^u_{t,x}(s),u(s)\big)\,ds\right| \\
&\le \int_t^T e^{\lambda(t-s)}\,\big|h\big(s,x^u_{t,x}(s),u(s)\big)\big|\,ds \\
&\le \int_t^T e^{\lambda(t-s)}\,M_h\,ds
\\ &= \int_0^{T-t} e^{-\lambda r}\,M_h\,dr \eqqcolon B(t).
\end{align*}
Therefore $-B(t)\le J_\lambda(t,x;u)\le B(t)$ for all $u\in\mathcal M(t)$, and taking the infimum over $u$ gives
\[
-B(t)\le V_\lambda(t,x)=\inf_{u\in\mathcal M(t)} J_\lambda(t,x;u)\le B(t).
\]
Thus $|V_\lambda(t,x)|\le B(t)$, and evaluating $B(t)$ yields \eqref{eq:bound_value}.
\end{pf}

\section{Lipschitz in state}\label{lem:Lip_x_appen}
\begin{lemmastar}[Lipschitz in state]
Given the above assumptions, for fixed $t$,
\begin{equation}
|V_\lambda(t,x_1)-V_\lambda(t,x_2)|
\ \le\ \Gamma_\lambda(t)\,\|x_1-x_2\|,
\label{eq:Lip_bound_V}
\end{equation}
where
\[\Gamma_\lambda(t):=L_h\int_0^{T-t} e^{(L_f-\lambda)r}\,dr,\]
\end{lemmastar}

\begin{pf}
Fix $t\in[0,T]$ and $x_1,x_2\in\mathbb R^n$. Let $u(\cdot)\in\mathcal M(t)$ be any admissible control.
Denote the corresponding trajectories by $x_i(s):=x^u_{t,x_i}(s)$ for $i\in\{1,2\}$.

\smallskip
\noindent\textit{Step 1: Trajectory sensitivity (Grönwall).}
By \assref{a:f_lipschitz}, for all $s\in[t,T]$,
\begin{multline*}
    \frac{d}{ds}\|x_1(s)-x_2(s)\|
\le \|f(x_1(s),u(s))-f(x_2(s),u(s))\| \\
\le L_f\|x_1(s)-x_2(s)\|
\end{multline*}
Hence, by Gr\"onwall's inequality,
\begin{equation}\label{eq:traj_lip}
\|x_1(s)-x_2(s)\|\le e^{L_f(s-t)}\|x_1-x_2\|,\quad s\in[t,T].
\end{equation}

\smallskip
\noindent\textit{Step 2: Cost difference under the same control.}
Using \assref{a:h_lipschitz}, the discount weight $e^{\lambda(t-s)}=e^{-\lambda(s-t)}$, and \eqref{eq:traj_lip},
\begin{multline*}
\big|J_\lambda(t,x_1;u)-J_\lambda(t,x_2;u)\big| \\
=\left|\int_t^T e^{\lambda(t-s)}\Big(h(s,x_1(s),u(s))-h(s,x_2(s),u(s))\Big)\,ds\right|\\
\le \int_t^T e^{-\lambda(s-t)}\,L_h\|x_1(s)-x_2(s)\|\,ds\\
\le L_h\int_t^T e^{-\lambda(s-t)}e^{L_f(s-t)}\,ds\;\|x_1-x_2\|\\
= L_h\int_0^{T-t} e^{(L_f-\lambda)r}\,dr\;\|x_1-x_2\|.
\end{multline*}
Define
\[
\Gamma_\lambda(t):=L_h\int_0^{T-t} e^{(L_f-\lambda)r}\,dr,
\]
so that
\begin{equation}\label{eq:J_lip}
\big|J_\lambda(t,x_1;u)-J_\lambda(t,x_2;u)\big|\le \Gamma_\lambda(t)\|x_1-x_2\|
\qquad \forall\,u\in\mathcal M(t).
\end{equation}

\smallskip
\noindent\textit{Step 3: Pass to the value function via $\varepsilon$-optimal controls.}
Fix $\varepsilon>0$ and choose $u_\varepsilon\in\mathcal M(t)$ such that
\[
J_\lambda(t,x_1;u_\varepsilon)\le V_\lambda(t,x_1)+\varepsilon.
\]
Then by \eqref{eq:J_lip},
\begin{multline*}
    V_\lambda(t,x_2)\le J_\lambda(t,x_2;u_\varepsilon)
\le J_\lambda(t,x_1;u_\varepsilon)+\Gamma_\lambda(t)\|x_1-x_2\| \\
\le V_\lambda(t,x_1)+\varepsilon+\Gamma_\lambda(t)\|x_1-x_2\|
\end{multline*}
Letting $\varepsilon\downarrow 0$ gives
\[
V_\lambda(t,x_2)-V_\lambda(t,x_1)\le \Gamma_\lambda(t)\|x_1-x_2\|.
\]
Interchanging the roles of $x_1$ and $x_2$ yields the reverse inequality, hence
\[
|V_\lambda(t,x_1)-V_\lambda(t,x_2)|\le \Gamma_\lambda(t)\|x_1-x_2\|.
\]

\smallskip
\noindent\textit{Step 4: Closed form.}
If $L_f\neq \lambda$ then
\[
\Gamma_\lambda(t)=L_h\int_0^{T-t} e^{(L_f-\lambda)r}\,dr
=\frac{L_h}{L_f-\lambda}\big(e^{(L_f-\lambda)(T-t)}-1\big),
\]
and if $L_f=\lambda$ then $\Gamma_\lambda(t)=L_h(T-t)$.
This proves \eqref{eq:Lip_bound_V}.
\end{pf}

\section{Local confinement from linear growth}\label{lem:local_confinement_appen}
\begin{lemmastar}[Local confinement from linear growth]
Under Assumption~\ref{a:f_uniform}, for every $x_0\in\mathbb R^n$ and every $\rho>0$, there exists $\delta_{\rho,x_0}>0$ such that, for every measurable control $u(\cdot)$ and every corresponding trajectory $y(\cdot)$ solving
\[
\dot y(r)=f(y(r),u(r)),\qquad y(0)=x_0,
\]
one has
\[
y(r)\in B_\rho(x_0)\qquad \forall r\in[0,\delta_{\rho,x_0}].
\]
Moreover, for every $r\ge 0$,
\begin{equation}\label{eq:local_growth_disp}
\|y(r)-x_0\| \le \big(e^{C_f r}-1\big)\,(\|x_0\|+1),
\end{equation}
where $C_f$ is the constant from Assumption~\ref{a:f_uniform}.
\end{lemmastar}

\begin{pf}
From Assumption~\ref{a:f_uniform},
\[
\|\dot y(r)\|=\|f(y(r),u(r))\|\le C_f(1+\|y(r)\|).
\]
Hence
\[
\|y(r)\|+1\le \|x_0\|+1 + C_f\int_0^r (\|y(s)\|+1)\,ds.
\]
By Gr\"onwall's inequality,
\[
\|y(r)\|+1\le e^{C_f r}(\|x_0\|+1).
\]
Therefore,
\begin{multline*}
    \|y(r)-x_0\|
\le \int_0^r \|f(y(s),u(s))\|\,ds\\
\le \int_0^r C_f(1+\|y(s)\|)\,ds 
\le \big(e^{C_f r}-1\big)(\|x_0\|+1)
\end{multline*}
which proves \eqref{eq:local_growth_disp}. Choosing $\delta_{\rho,x_0}>0$ so that
\[
\big(e^{C_f \delta_{\rho,x_0}}-1\big)(\|x_0\|+1)<\rho
\]
gives the local confinement claim.
\end{pf}

\section{Lipschitz in time}\label{lem:time_cont_appen}
\begin{lemmastar}[Time continuity]
Under \assref{a:h_uniform}, \assref{a:f_uniform}, and \lemref{lem:Lip_x}, for $\sigma\in[0,T-t]$,
\begin{multline}\label{eq:time_cont}
    |V_\lambda(t+\sigma,x)-V_\lambda(t,x)|
\ \le\ M_h\!\int_{0}^{\sigma}\! e^{-\lambda r}dr\\
+\,e^{-\lambda\sigma}\Gamma_\lambda(t+\sigma)\big(e^{C_f\sigma}-1\big)(\|x\|+1)
\, \\ +\,|1-e^{-\lambda\sigma}|\,\|V_\lambda\|_\infty
\end{multline}.
\end{lemmastar}

\begin{pf}
Fix $(t,x)\in[0,T]\times\mathbb R^n$ and $\sigma\in[0,T-t]$. By the discounted DPP
(Lemma~\ref{lem:DPP_rel_fnl}),
\begin{align}
V_\lambda(t,x)
=\inf_{u\in\mathcal M(t)}\Big\{
I_\sigma(t,x;u)+e^{-\lambda\sigma}V_\lambda\big(t+\sigma, X_u\big)
\Big\},
\label{eq:DPP_used}
\end{align}
where
\begin{multline*}
I_\sigma(t,x;u):=\int_t^{t+\sigma} e^{\lambda(t-s)}h\big(s,x^u_{t,x}(s),u(s)\big)\,ds,
\\
X_u:=x^u_{t,x}(t+\sigma).
\end{multline*}

\noindent\textit{Step 1: bound the head integral.}
By \assref{a:h_uniform},
\begin{multline}
|I_\sigma(t,x;u)|
\le \int_t^{t+\sigma} e^{\lambda(t-s)}M_h\,ds \\
= M_h\int_0^\sigma e^{-\lambda r}\,dr
\qquad \forall\,u\in\mathcal M(t).
\label{eq:head_bound_new}
\end{multline}

\noindent\textit{Step 2: bound the state displacement at time $t+\sigma$.}
By Lemma~\ref{lem:local_confinement_appen},
\begin{equation}
\|X_u-x\|
\le \big(e^{C_f\sigma}-1\big)(\|x\|+1).
\label{eq:state_drift_new}
\end{equation}

\noindent\textit{Step 3: compare $V_\lambda(t+\sigma,X_u)$ to $V_\lambda(t+\sigma,x)$.}
By Lemma~\ref{lem:Lip_x} at time $t+\sigma$,
\begin{multline}
\big|V_\lambda(t+\sigma,X_u)-V_\lambda(t+\sigma,x)\big| \\
\le \Gamma_\lambda(t+\sigma)\,\|X_u-x\|
\le \Gamma_\lambda(t+\sigma)\big(e^{C_f\sigma}-1\big)(\|x\|+1).
\label{eq:Lip_at_tps_new}
\end{multline}

\noindent\textit{Step 4: sandwich $V_\lambda(t,x)$ around $e^{-\lambda\sigma}V_\lambda(t+\sigma,x)$.}
Combining \eqref{eq:DPP_used}, \eqref{eq:head_bound_new}, and \eqref{eq:Lip_at_tps_new} yields
\begin{multline}
\big|V_\lambda(t,x)-e^{-\lambda\sigma}V_\lambda(t+\sigma,x)\big| \\
\le M_h\int_0^\sigma e^{-\lambda r}dr
+ e^{-\lambda\sigma}\Gamma_\lambda(t+\sigma)\big(e^{C_f\sigma}-1\big)(\|x\|+1).
\label{eq:key_mid_bound_new}
\end{multline}

\noindent\textit{Step 5: remove the discount mismatch.}
Using the triangle inequality,
\begin{multline*}
|V_\lambda(t+\sigma,x)-V_\lambda(t,x)| \\
\le |1-e^{-\lambda\sigma}|\,\|V_\lambda\|_\infty
+ M_h\int_0^\sigma e^{-\lambda r}dr \\
+ e^{-\lambda\sigma}\Gamma_\lambda(t+\sigma)\big(e^{C_f\sigma}-1\big)(\|x\|+1),
\end{multline*}
which is exactly \eqref{eq:time_cont}.
\end{pf}

\end{proofver}

\section{Local existence}\label{lem:local_exist_appen}
\begin{lemmastar}
Assume $h$ is uniformly continuous and
\[
\phi_t+H(t_0,x_0,D\phi)-\lambda\phi\ \le\ -\theta
\quad (\theta>0).
\]
Then $\exists\,u^\ast\in\mathcal U,\ \delta_0>0$ such that, for
$x$ solving $\dot x=f(x,u^\ast)$, $x(t_0)=x_0$, and all
$\delta\in(0,\delta_0]$,
\begin{align}
&e^{-\lambda\delta}\phi(t_0+\delta,x(\delta))-\phi(t_0,x_0)
\notag\\[-2pt]
&\quad + \int_{0}^{\delta}\! e^{-\lambda r}\,
         h(t_0+r,x(r),u^\ast)\,dr
\ \le\ -\frac{\theta}{2}\!\int_{0}^{\delta}\! e^{-\lambda r}dr.
\label{eq:local_exist}
\end{align}
\end{lemmastar}

\begin{pf}
Let $p_0:=D\phi(t_0,x_0)$. The assumption
\[
\phi_t(t_0,x_0)+H(t_0,x_0,p_0)-\lambda\phi(t_0,x_0)\le -\theta
\]
means
\begin{multline*}
    \inf_{u\in\mathcal U}\Big\{\phi_t(t_0,x_0)+p_0\cdot f(x_0,u) \\+h(t_0,x_0,u)-\lambda\phi(t_0,x_0)\Big\}\le -\theta
\end{multline*}
By compactness of $\mathcal U$ and continuity in $u$ of the minimized expression,
there exists $u^\ast\in\mathcal U$ such that
\[
\Lambda_\lambda(t_0,x_0,u^\ast;\phi)\le -\tfrac{3}{4}\theta.
\]

By continuity of $\Lambda_\lambda(\cdot,\cdot,u^\ast;\phi)$ in $(s,x)$ at $(t_0,x_0)$,
there exists a neighborhood and $\delta_0>0$ such that
\begin{multline*}
    \Lambda_\lambda(t_0+r,\,y,\,u^\ast;\phi)\le -\tfrac{1}{2}\theta \\
\quad \forall r\in[0,\delta_0],\ \forall y \text{ with }\|y-x_0\|\le \rho
\end{multline*}
for some $\rho>0$.

Let $y(\cdot)$ solve the shifted ODE
\[
\dot y(r)=f(y(r),u^\ast),\qquad y(0)=x_0.
\]
By continuity of trajectories, shrinking $\delta_0$ if needed we ensure
$y(r)\in B_\rho(x_0)$ for all $r\in[0,\delta]$ whenever $\delta\in(0,\delta_0]$.
Hence, for all such $\delta$,
\[
\Lambda_\lambda(t_0+r,\,y(r),\,u^\ast;\phi)\le -\tfrac{1}{2}\theta
\qquad \forall r\in[0,\delta].
\]

Now define $g(r):=e^{-\lambda r}\phi(t_0+r,y(r))$. By the chain rule,
\[
g'(r)=e^{-\lambda r}\big(\phi_t + D\phi\cdot f -\lambda\phi\big)(t_0+r,y(r),u^\ast).
\]
Therefore,
\begin{multline*}
    e^{-\lambda\delta}\phi(t_0+\delta,y(\delta))-\phi(t_0,x_0) \\
+\int_0^\delta e^{-\lambda r}h(t_0+r,y(r),u^\ast)\,dr\\
= \int_0^\delta e^{-\lambda r}\Lambda_\lambda(t_0+r,y(r),u^\ast;\phi)\,dr 
\ \le\ -\tfrac{\theta}{2}\int_0^\delta e^{-\lambda r}\,dr
\end{multline*}
which is \eqref{eq:local_exist}.
\end{pf}


\section{Universal existence}\label{lem:local_univ_appen}
\begin{lemmastar}
Assume $h$ is uniformity continuous and
\[
\phi_t+H(t_0,x_0,D\phi)-\lambda\phi\ \ge\ \theta>0.
\]
Then $\exists\,\delta_0>0$ such that, for every measurable $u(\cdot)$
and the trajectory $x(\cdot)$ on $[t_0,t_0+\delta]$,
\begin{align}
&e^{-\lambda\delta}\phi(t_0+\delta,x(\delta))-\phi(t_0,x_0)
\notag\\[-2pt]
&\quad + \int_{0}^{\delta}\! e^{-\lambda r}\,
         h(t_0+r,x(r),u(r))\,dr
\ \ge\ \frac{\theta}{2}\!\int_{0}^{\delta}\! e^{-\lambda r}dr.
\label{eq:local_univ}
\end{align}
\end{lemmastar}

\begin{pf}
Let $p_0:=D\phi(t_0,x_0)$. Define
The assumption
\[
\phi_t(t_0,x_0)+H(t_0,x_0,p_0)-\lambda\phi(t_0,x_0)\ge \theta
\]
means
\[
\inf_{u\in\mathcal U}\Lambda_\lambda(t_0,x_0,u;\phi)\ge \theta,
\]
hence
\begin{equation}\label{eq:Lambda_pointwise}
\Lambda_\lambda(t_0,x_0,u;\phi)\ge \theta \qquad \forall\,u\in\mathcal U.
\end{equation}

By continuity of $(s,x,u)\mapsto \Lambda_\lambda(s,x,u;\phi)$ and compactness of $\mathcal U$,
the lower bound \eqref{eq:Lambda_pointwise} is uniform: there exist $\rho>0$ and $\delta_0>0$
such that
\begin{multline}\label{eq:Lambda_uniform_nbhd}
\Lambda_\lambda(t_0+r,y,u;\phi)\ge \tfrac{\theta}{2} \\
\forall r\in[0,\delta_0],\ \forall y\in B_\rho(x_0),\ \forall u\in\mathcal U.
\end{multline}

Now fix any measurable control $u(\cdot)$ on $[0,\delta]$ and let $y(\cdot)$ solve the shifted ODE
\[
\dot y(r)=f(y(r),u(r)),\qquad y(0)=x_0.
\]
\begin{proofver}
By Lemma~\ref{lem:local_confinement_appen}, after shrinking $\delta_0$ if necessary, we may assume that every trajectory starting from $x_0$ remains in $B_\rho(x_0)$ on $[0,\delta]$ for all $\delta\in(0,\delta_0]$, uniformly over all measurable controls $u(\cdot)$. Hence \eqref{eq:Lambda_uniform_nbhd} gives
\[
\Lambda_\lambda(t_0+r,y(r),u(r);\phi)\ge \tfrac{\theta}{2}\qquad \forall r\in[0,\delta].
\]
\end{proofver}
\begin{ideaver}
    After shrinking $\delta_0$ if necessary, we may assume that every trajectory starting from $x_0$ remains in $B_\rho(x_0)$ on $[0,\delta]$ for all $\delta\in(0,\delta_0]$, uniformly over all measurable controls $u(\cdot)$. Hence \eqref{eq:Lambda_uniform_nbhd} gives
\[
\Lambda_\lambda(t_0+r,y(r),u(r);\phi)\ge \tfrac{\theta}{2}\qquad \forall r\in[0,\delta].
\]
\end{ideaver}

Define $g(r):=e^{-\lambda r}\phi(t_0+r,y(r))$. By the chain rule,
\[
g'(r)=e^{-\lambda r}\big(\phi_t+D\phi\cdot f-\lambda\phi\big)(t_0+r,y(r),u(r)).
\]
Therefore,
\begin{multline*}
    e^{-\lambda\delta}\phi(t_0+\delta,y(\delta))-\phi(t_0,x_0) \\
+\int_0^\delta e^{-\lambda r}h(t_0+r,y(r),u(r))\,dr\\
= \int_0^\delta e^{-\lambda r}\Lambda_\lambda(t_0+r,y(r),u(r);\phi)\,dr
\ \ge\ \tfrac{\theta}{2}\int_0^\delta e^{-\lambda r}\,dr
\end{multline*}
which is \eqref{eq:local_univ}.
\end{pf}

\section{Viscosity characterization}\label{thm:visc_V_appen}

\begin{proofver}
\begin{theoremstar}[Viscosity characterization]
Under \assref{a:U_compact}-\assref{a:h_continuity_u},
$V_\lambda$ is a bounded, continuous and unique viscosity solution of
\begin{multline}
V_{\lambda,t}(t,x)+H\!\big(t,x,\nabla_x V_\lambda(t,x)\big)
-\lambda\,V_\lambda(t,x)=0, \\
V_\lambda(T,x)=0
\label{eq:HJB_rel}
\end{multline}
\end{theoremstar}
\end{proofver}

\begin{ideaver}
    The Proof of Theorem \ref{thm:visc_V}
\end{ideaver}

\begin{pf}
We prove the viscosity sub- and super-solution inequalities on $[0,T)\times\mathbb R^n$
and note that the terminal condition $V_\lambda(T,x)=0$ holds by definition.

\smallskip
\noindent\textbf{(i) Subsolution.}
Let $\phi\in C^1$ and suppose $V_\lambda-\phi$ has a local maximum at $(t_0,x_0)$ with $t_0<T$.
without loss of generality assume $(V_\lambda-\phi)(t_0,x_0)=0$, i.e.\ $\phi(t_0,x_0)=V_\lambda(t_0,x_0)$.
By the definition of local maximum and continuity of $V_\lambda-\phi$, for every $\varepsilon>0$
there exist $\rho>0$ and $\delta_1>0$ such that
\begin{equation}\label{eq:touch_max_eps}
-\varepsilon \ \le\ (V_\lambda-\phi)(t_0+r,y)\ \le\ 0
\quad \forall r\in[0,\delta_1],\ \forall y\in B_\rho(x_0).
\end{equation}

\begin{proofver}
By Lemma~\ref{lem:local_confinement_appen}, for the given $\rho>0$ there exists $\delta_{\rho,x_0}>0$ such that every trajectory starting from $x_0$ remains in $B_\rho(x_0)$ on $[0,\delta]$ whenever $\delta\in(0,\delta_{\rho,x_0}]$, uniformly over all measurable controls.
Choose $\delta\in(0,\min\{\delta_1,\delta_{\rho,x_0}\}]$.
\end{proofver}

\begin{ideaver}
    for the given $\rho>0$ there exists $\delta_{\rho,x_0}>0$ such that every trajectory starting from $x_0$ remains in $B_\rho(x_0)$ on $[0,\delta]$ whenever $\delta\in(0,\delta_{\rho,x_0}]$, uniformly over all measurable controls.
Choose $\delta\in(0,\min\{\delta_1,\delta_{\rho,x_0}\}]$.
\end{ideaver}

We need to prove that 
\[
\phi_t(t_0,x_0)+H(t_0,x_0,D\phi(t_0,x_0))-\lambda\,\phi(t_0,x_0)\ \ge\ 0.
\]
Suppose, for contradiction, that there exists $\theta>0$ such that
\begin{equation}\label{eq:sub_contra}
\phi_t(t_0,x_0)+H(t_0,x_0,D\phi(t_0,x_0))-\lambda\,\phi(t_0,x_0)\ \le\ -\theta.
\end{equation}
By Lemma~\ref{lem:local_exist_appen}, there exist a control $u^\ast\in\mathcal U$
and $\delta_0>0$ such that, for all $\delta\in(0,\min\{\delta_0,\delta_1,\delta_{\rho,x_0}\}]$,
the associated shifted trajectory $y(\cdot)$ satisfies

\begin{multline*}
\int_0^\delta e^{-\lambda r}h(t_0+r,y(r),u(r))\,dr
+e^{-\lambda\delta}\phi(t_0+\delta,y(\delta)) \\
\ \le\ \phi(t_0,x_0)-\frac{\theta}{2}\int_0^\delta e^{-\lambda r}\,dr.
\end{multline*}
Thus
\begin{multline}\label{eq:local_exist_use}
\inf_{u\in \mathcal{U}} \Big\{ \int_0^\delta e^{-\lambda r}h(t_0+r,y(r),u^\ast)\,dr
+ \\ e^{-\lambda\delta}\phi(t_0+\delta,y(\delta))
\ - \ \phi(t_0,x_0)\Big\}\le-\frac{\theta}{2}\int_0^\delta e^{-\lambda r}\,dr.
\end{multline}

On the other hand, \eqref{eq:touch_max_eps} implies
\begin{multline*}
    e^{-\lambda\delta}V_\lambda(t_0+\delta,y(\delta))- e^{-\lambda\delta}\phi(t_0+\delta,y(\delta)) \\ \le V_\lambda(t_0,x_0)\ - \phi(t_0,x_0)
\end{multline*}
Combining with \eqref{eq:local_exist_use} yields, that there exists a $u^\ast$,
\begin{align*}
&\int_0^\delta e^{-\lambda r}h(t_0+r,y(r),u^\ast)\,dr
+e^{-\lambda\delta}V_\lambda(t_0+\delta,y(\delta)) \\
&\qquad\le\ V_\lambda(t_0,x_0)-\frac{\theta}{2}\int_0^\delta e^{-\lambda r}\,dr 
\end{align*}

Taking the infimum over $u(\cdot)$ and using the shifted DPP
(Lemma~\ref{lem:DPP_rel_fnl} written on $[t_0,t_0+\delta]$) gives
\[
V_\lambda(t_0,x_0)\ \ge\ V_\lambda(t_0,x_0)+\frac{\theta}{2}\int_0^\delta e^{-\lambda r}\,dr
\]
Since $\theta>0$, this yields a contradiction.
Hence \eqref{eq:sub_contra} is false, proving the subsolution inequality:
\[
\phi_t(t_0,x_0)+H(t_0,x_0,D\phi(t_0,x_0))-\lambda\,V_\lambda(t_0,x_0)\ \ge\ 0.
\]

\smallskip
\noindent\textbf{(ii) Supersolution.}
Let $\phi\in C^1$ and suppose $V_\lambda-\phi$ has a local minimum at $(t_0,x_0)$ with $t_0<T$.
Again normalize $(V_\lambda-\phi)(t_0,x_0)=0$.
Then for every $\varepsilon>0$ there exist $\rho>0$ and $\delta_1>0$ such that
\begin{equation}\label{eq:touch_min_eps}
0 \ \le\ (V_\lambda-\phi)(t_0+r,y)\ \le\ \varepsilon
\quad \forall r\in[0,\delta_1],\ \forall y\in B_\rho(x_0).
\end{equation}
In particular, $V_\lambda(t_0+\delta,y)\le \phi(t_0+\delta,y)+\varepsilon$ on this neighborhood.

We claim that
\[
\phi_t(t_0,x_0)+H(t_0,x_0,D\phi(t_0,x_0))-\lambda\,\phi(t_0,x_0)\ \le\ 0.
\]
Suppose, for contradiction, that there exists $\theta>0$ such that
\begin{equation}\label{eq:super_contra}
\phi_t(t_0,x_0)+H(t_0,x_0,D\phi(t_0,x_0))-\lambda\,\phi(t_0,x_0)\ \ge\ \theta.
\end{equation}
By Lemma~\ref{lem:local_univ_appen}, for every measurable $u(\cdot) \in\mathcal U$
,there exists $ \delta_0>0$ such that, for all $\delta\in(0,\min\{\delta_0,\delta_1,\delta_{\rho,x_0}\}]$,
the associated shifted trajectory $y(\cdot)$ satisfies
\begin{multline}\label{eq:local_univ_use}
\int_0^\delta e^{-\lambda r}h(t_0+r,y(r),u(\cdot))\,dr
+e^{-\lambda\delta}\phi(t_0+\delta,y(\delta)) \\
\ \ge\ \phi(t_0,x_0)-\frac{\theta}{2}\int_0^\delta e^{-\lambda r}\,dr.
\end{multline}
On the other hand, \eqref{eq:touch_min_eps} implies
\begin{multline*}
    e^{-\lambda\delta}V_\lambda(t_0+\delta,y(\delta))- e^{-\lambda\delta}\phi(t_0+\delta,y(\delta)) \\ \ge V_\lambda(t_0,x_0)\ - \phi(t_0,x_0)
\end{multline*}
Combining with \eqref{eq:local_univ_use} yields, that for all $u(\cdot) \in \mathcal U$,
\begin{align*}
&\int_0^\delta e^{-\lambda r}h(t_0+r,y(r),u(\cdot))\,dr
+e^{-\lambda\delta}V_\lambda(t_0+\delta,y(\delta)) \\
&\qquad\ge\ V_\lambda(t_0,x_0)-\frac{\theta}{2}\int_0^\delta e^{-\lambda r}\,dr 
\end{align*}

Using the DPP and taking the infimum over controls gives
\[
V_\lambda(t_0,x_0)\ \le\ V_\lambda(t_0,x_0) - \frac{\theta}{2}\int_0^\delta e^{-\lambda r}\,dr
\]
Since $\theta>0$, this yields a contradiction.
Thus \eqref{eq:super_contra} is false and we conclude the supersolution inequality:
\[
\phi_t(t_0,x_0)+H(t_0,x_0,D\phi(t_0,x_0))-\lambda\,V_\lambda(t_0,x_0)\ \le\ 0.
\]
\begin{proofver}
\smallskip
\noindent\textbf{(iii) Conclusion.}
Parts (i) and (ii) show that $V_\lambda$ is a viscosity solution of \eqref{eq:HJB_rel} on $[0,T)\times\mathbb R^n$.
The terminal condition $V_\lambda(T,x)=0$ holds by definition.
Uniqueness among bounded continuous viscosity solutions follows from the comparison principle
for proper Hamilton--Jacobi equations (for $\lambda>0$, the term $-\lambda V$ makes the PDE strictly proper).
\end{pf}
\end{proofver}
\begin{ideaver}
\end{pf}
\end{ideaver}

\begin{proofver}
\section{Monotonicity and stability}\label{lem:num_mono_stable_appen}
\begin{lemma}[Monotonicity and stability]
For bounded $\Psi_1\le\Psi_2$, one has
$\widehat{\mathcal T}_{\sigma,\lambda}\Psi_1\le \widehat{\mathcal T}_{\sigma,\lambda}\Psi_2$ (monotone).
Moreover, if $|h|\le M_h$ and $\widehat c_{\sigma,\lambda}$ is bounded accordingly, then
$\widehat{\mathcal T}_{\sigma,\lambda}$ maps bounded functions to bounded functions (stability).
\end{lemma}

\begin{pf}
Monotonicity is immediate from \eqnref{eq:num_Bellman_def} since $\Psi$ appears only inside
$e^{-\lambda\sigma}\Psi(\cdot)$ with a positive coefficient. Stability follows by bounding
$\widehat c_{\sigma,\lambda}$ using $|h|\le M_h$ and taking $\sup$ over $(\tau,x)$.
\end{pf}

\section{Consistency}\label{lem:num_consistency_appen}
\begin{lemma}[Consistency]
Let $\phi\in C^1([0,T]\times\R^n)$ with bounded derivatives. Then
\begin{multline}
\frac{(\widehat{\mathcal T}_{\sigma,\lambda}\phi)(\tau,x)-\phi(\tau,x)}{\sigma}
\ \xrightarrow[\ \sigma\downarrow 0\ ]{}\ \\
-\,\phi_\tau(\tau,x)+\widetilde H(\tau,x,\nabla_x\phi(\tau,x))
-\lambda\,\phi(\tau,x),
\label{eq:num_consistency_limit_appen}
\end{multline}
uniformly on compact subsets of $(0,T]\times\R^n$.
\end{lemma}

\begin{pf}
Fix $(\tau,x)$ and $u\in\mathcal U$. Using \eqnref{eq:num_Bellman_def},
\[
(\widehat{\mathcal T}_{\sigma,\lambda}\phi)(\tau,x)
\le \widehat c_{\sigma,\lambda}(\tau,x,u)
+ e^{-\lambda\sigma}\phi\big(\tau-\sigma,\widehat F_\sigma(x,u)\big).
\]
Apply Taylor expansion of $\phi$ at $(\tau,x)$ and the consistency
\eqnref{eq:num_flow_consistency}-\eqnref{eq:num_cost_consistency}:
\begin{multline*}
\phi(\tau-\sigma,\widehat F_\sigma(x,u)) 
= \\ \phi(\tau,x) - \sigma \phi_\tau(\tau,x)
+ \nabla\phi(\tau,x)\cdot(\widehat F_\sigma(x,u)-x) + o(\sigma)
\\
= \phi(\tau,x) + \sigma\big(\nabla\phi\cdot f - \phi_\tau\big)(\tau,x) + o(\sigma),
\end{multline*}
and $e^{-\lambda\sigma}=1-\lambda\sigma+o(\sigma)$. Therefore,
\begin{multline*}
    (\widehat{\mathcal T}_{\sigma,\lambda}\phi)(\tau,x)-\phi(\tau,x) \\
\le \sigma\Big(h(T-\tau,x,u)+\nabla\phi\cdot f - \phi_\tau - \lambda\phi\Big)(\tau,x)
+ o(\sigma)
\end{multline*}
Divide by $\sigma$ and infimize over $u\in\mathcal U$ to get the $\limsup$ bound.
The matching $\liminf$ follows from the same expansion applied to a minimizing sequence
$u_\sigma$ (compactness of $\mathcal U$ and uniformity of the $o(\sigma)$ terms on compacts).
\end{pf}
\end{proofver}

\section{Sample-based RL training loop}
\label{appen:rl_training_loop}

\begin{algorithm}[H]
\caption{Sample-based training for discounted travel-cost value learning}
\label{alg:sample_based_value_learning}
\begin{algorithmic}[1]
\State \textbf{Input:} value network $W_\theta$, target network $\bar W_{\bar\theta}$
\State \textbf{Input:} discount rate $\lambda$, step size $\Delta\tau$, action set $\mathcal U$
\State \textbf{Input:} training state distribution $\mu$ over the ROI
\State $\gamma \gets e^{-\lambda \Delta\tau}$, \quad $w \gets (1-\gamma)/\lambda$
\For{$k=1,\dots,K$}
    \State sample a batch of states $\{x_i\}_{i=1}^B \sim \mu$
    \For{each $x_i$ in the batch}
        \For{each $u\in\mathcal U$}
            \State simulate one step from $x_i$ under control $u$
            \[
            x_i^+ \gets \mathrm{SimStep}(x_i,u,\Delta\tau)
            \]
            \State compute one-step cost
            \[
            c_i(x_i,u) \gets w\,h(x_i,u)
            \]
            \State form Bellman candidate
            \[
            y_i(u) \gets c_i(x_i,u)+\gamma\,\bar W_{\bar\theta}(x_i^+)
            \]
        \EndFor
        \State set TD target
        \[
        y_i \gets \min_{u\in\mathcal U} y_i(u)
        \]
    \EndFor
    \State update $\theta$ by minimizing
    \[
    \mathcal L(\theta)=\frac{1}{B}\sum_{i=1}^B \ell\!\left(W_\theta(x_i),\,y_i\right)
    \]
    \State update target network:
    \[
    \bar\theta \gets \tau \theta +(1-\tau)\bar\theta
    \]
\EndFor
\end{algorithmic}
\end{algorithm}

\section{Sensitivity to the discount parameter \texorpdfstring{$\lambda$}{lambda} on the 2D double integrator}
\label{app:lambda_sensitivity}

\begin{figure}[t]
    \centering
    \includegraphics[width=0.95\linewidth]{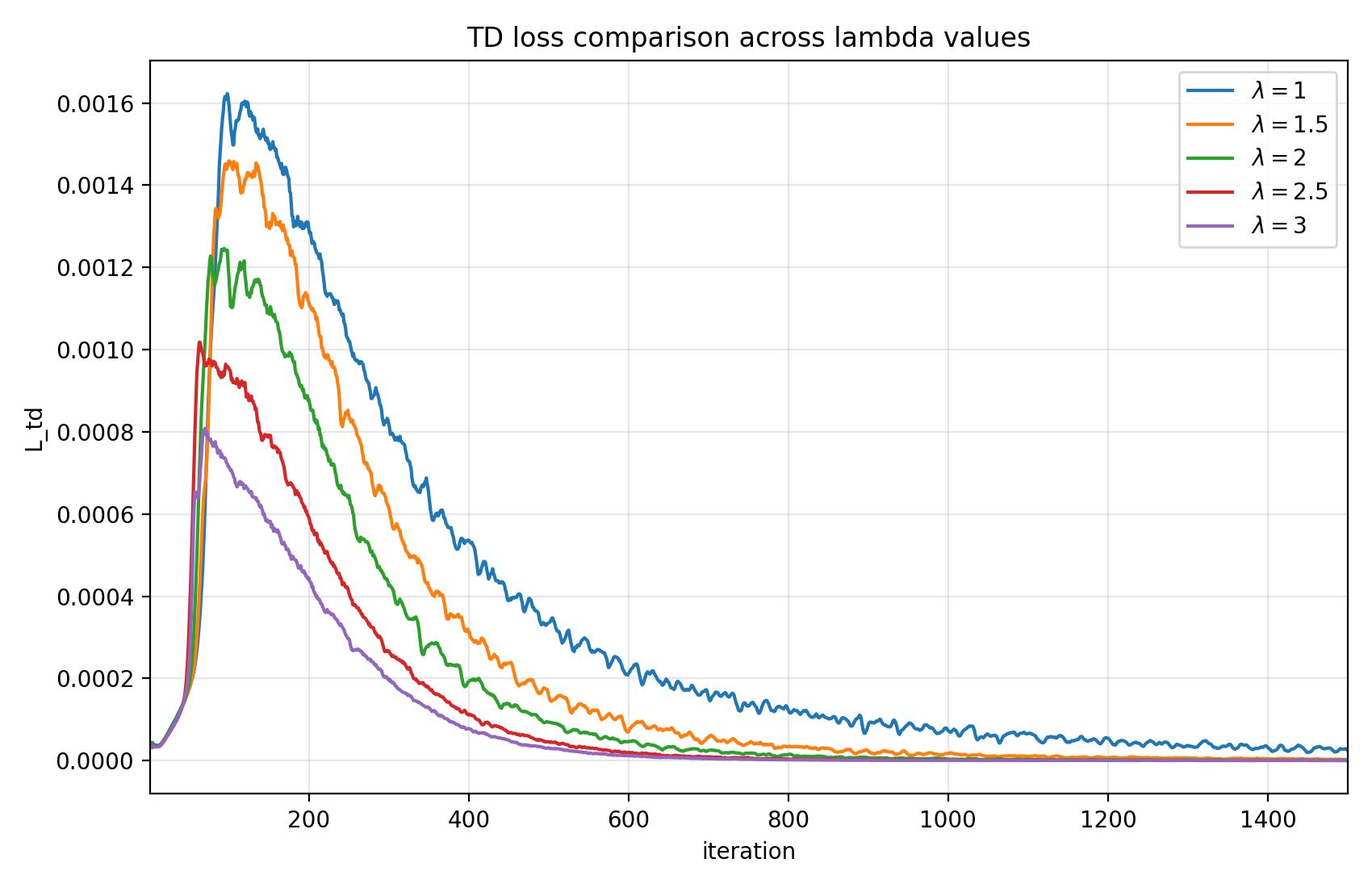}
    \caption{TD-loss comparison across discount parameters \(\lambda\) on the 2D double-integrator benchmark. Larger \(\lambda\) leads to faster TD-loss decay, consistent with the stronger contraction factor \(e^{-\lambda\Delta\tau}\).}
    \label{fig:lambda_td_compare}
\end{figure}

To study the practical role of the relative exponential discount, we repeated the 2D double-integrator experiment for several values of \(\lambda\), while keeping the remaining training and discretization settings fixed. Figure~\ref{fig:lambda_td_compare} compares the TD-loss trajectories for \(\lambda\in\{1,1.5,2,2.5,3\}\).

As \(\lambda\) increases, the TD loss decays more rapidly and reaches a lower level within the same iteration budget. This trend is consistent with the theoretical contraction factor \(e^{-\lambda\Delta\tau}\): larger \(\lambda\) yields a stronger one-step contraction and therefore a better-conditioned Bellman update. At the same time, changing \(\lambda\) also changes the discounted objective itself, so this experiment should be interpreted as a sensitivity study rather than a same-objective speed benchmark.

\section{Higher-dimensional demonstrations: 4D planar double integrator and 6D chained integrator}
\label{app:higher_dimensional_results}

To assess whether the Bellman/HJB connection remains visible beyond the 2D benchmark, we also trained value networks on a 4D planar double integrator and a 6D planar chained integrator. In these experiments, training was performed using TD loss. During training, we additionally evaluated a numerical HJB loss, but this quantity was not used for optimization.

\textbf{4D planar double integrator.}
The 4D system has state
\[
x=(p_x,p_y,v_x,v_y)\in\mathbb{R}^4,
\]
with dynamics
\begin{equation}
\dot p_x = v_x,\qquad
\dot p_y = v_y,\qquad
\dot v_x = u_x,\qquad
\dot v_y = u_y,
\label{eq:appendix_4d_dynamics}
\end{equation}
where the control is
\[
u=(u_x,u_y)\in[-a_{\max},a_{\max}]^2.
\]

\textbf{6D chained integrator.}
The 6D system has state
\[
x=(p_x,p_y,v_x,v_y,a_x,a_y)\in\mathbb{R}^6,
\]
with dynamics
\begin{multline}
\dot p_x = v_x,\qquad
\dot p_y = v_y,\qquad
\dot v_x = a_x,\\
\dot v_y = a_y,\qquad
\dot a_x = j_x,\qquad
\dot a_y = j_y,
\label{eq:appendix_6d_dynamics}
\end{multline}
where the control is
\[
u=(j_x,j_y)\in[-j_{\max},j_{\max}]^2.
\]

\textbf{Running cost.}
In both systems, the running cost depends only on planar position and is given by
\begin{equation}
h(x,u)=\alpha\,\min\!\bigl(0,\sqrt{p_x^2+p_y^2}-r\bigr),
\label{eq:appendix_hd_running_cost}
\end{equation}
so it is strictly negative inside the target disk of radius \(r\) and zero on and outside the disk.

\begin{figure}[H]
    \centering
    \includegraphics[width=0.8\linewidth]{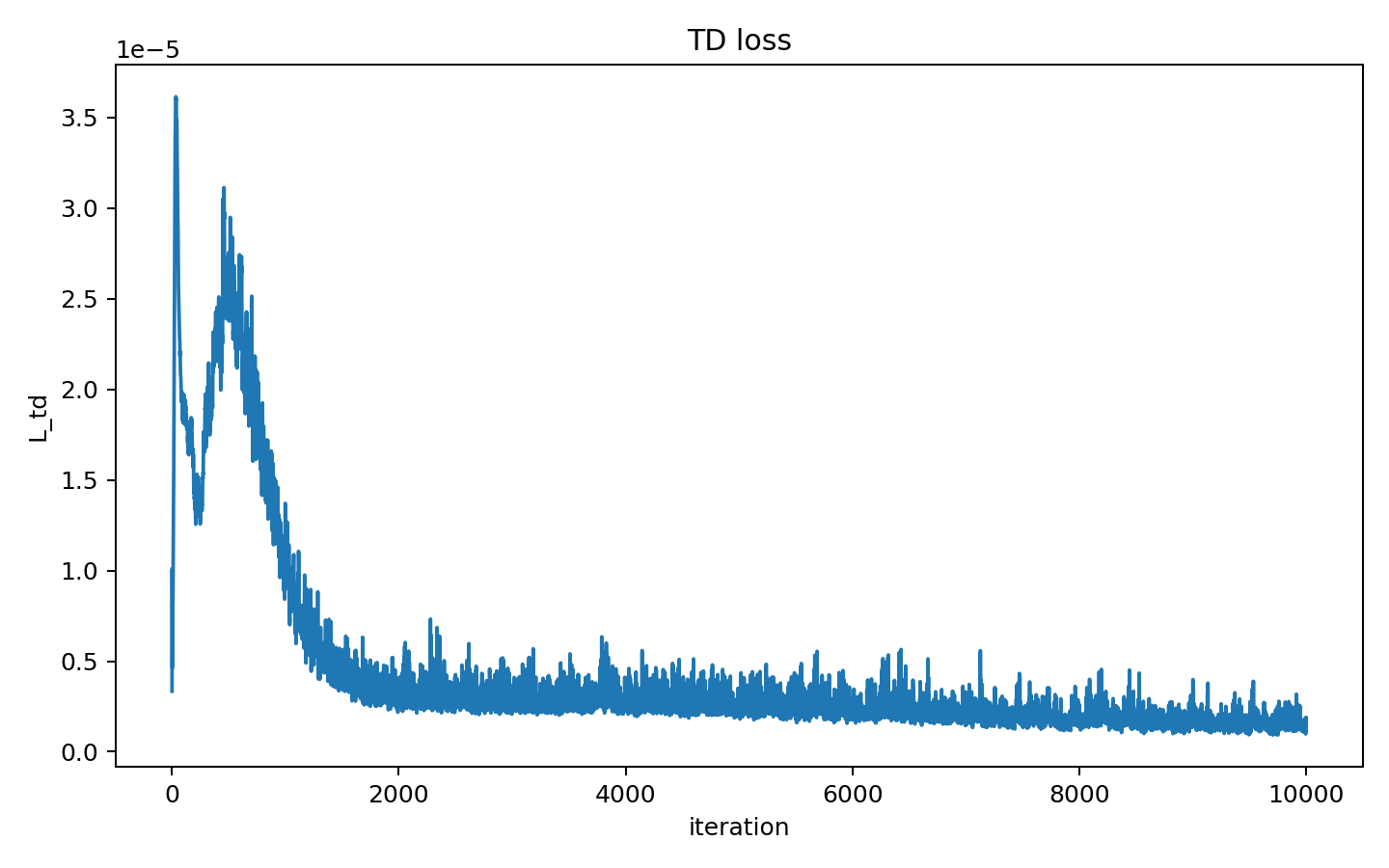}
    \caption{TD-loss trajectory for the 4D planar double-integrator experiment.}
    \label{fig:td_4d}
\end{figure}

\begin{figure}[H]
    \centering
    \includegraphics[width=0.8\linewidth]{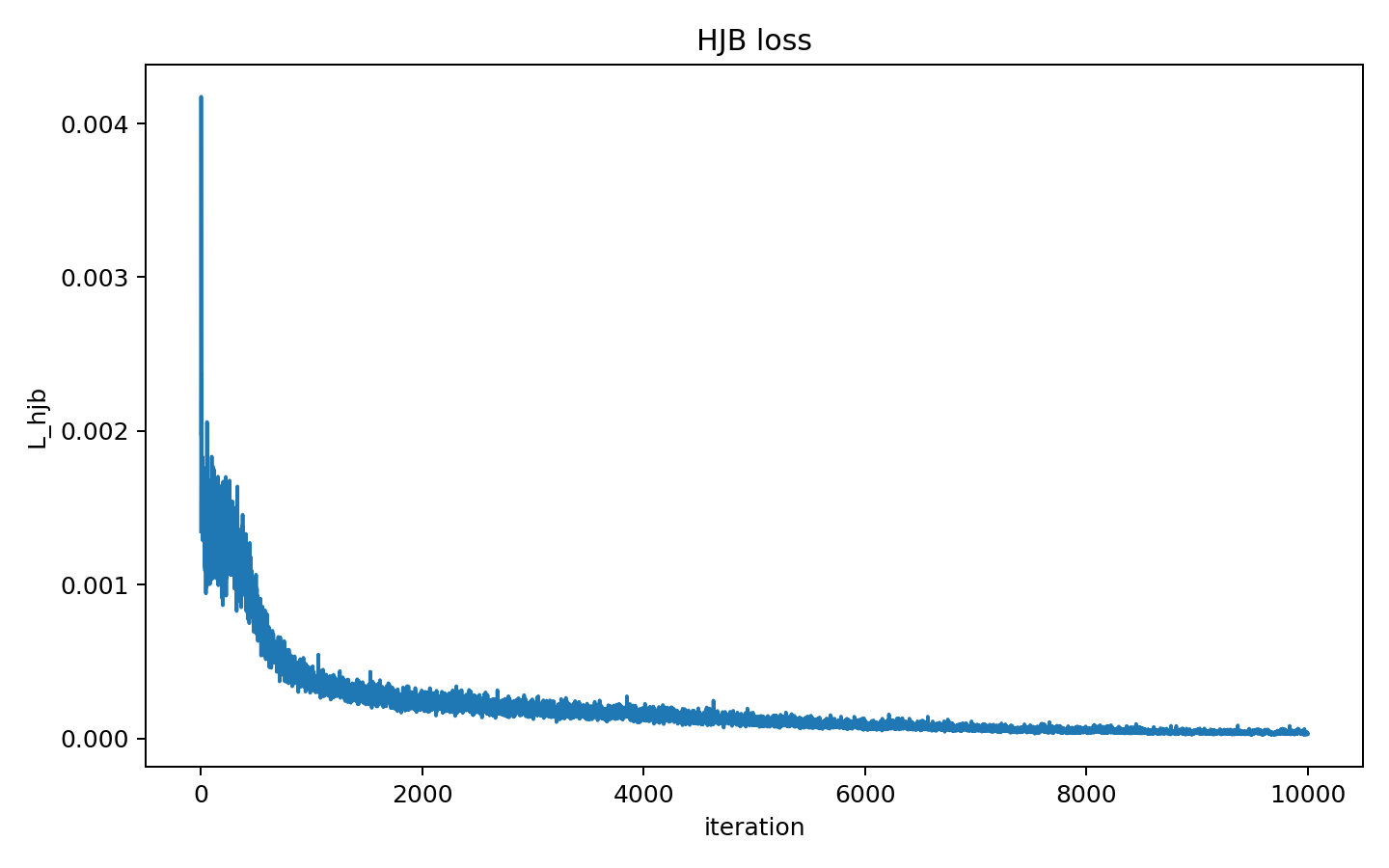}
    \caption{Numerically evaluated HJB loss during TD-only training for the 4D planar double-integrator experiment.}
    \label{fig:hjb_4d}
\end{figure}

\begin{figure}[H]
    \centering
    \includegraphics[width=0.8\linewidth]{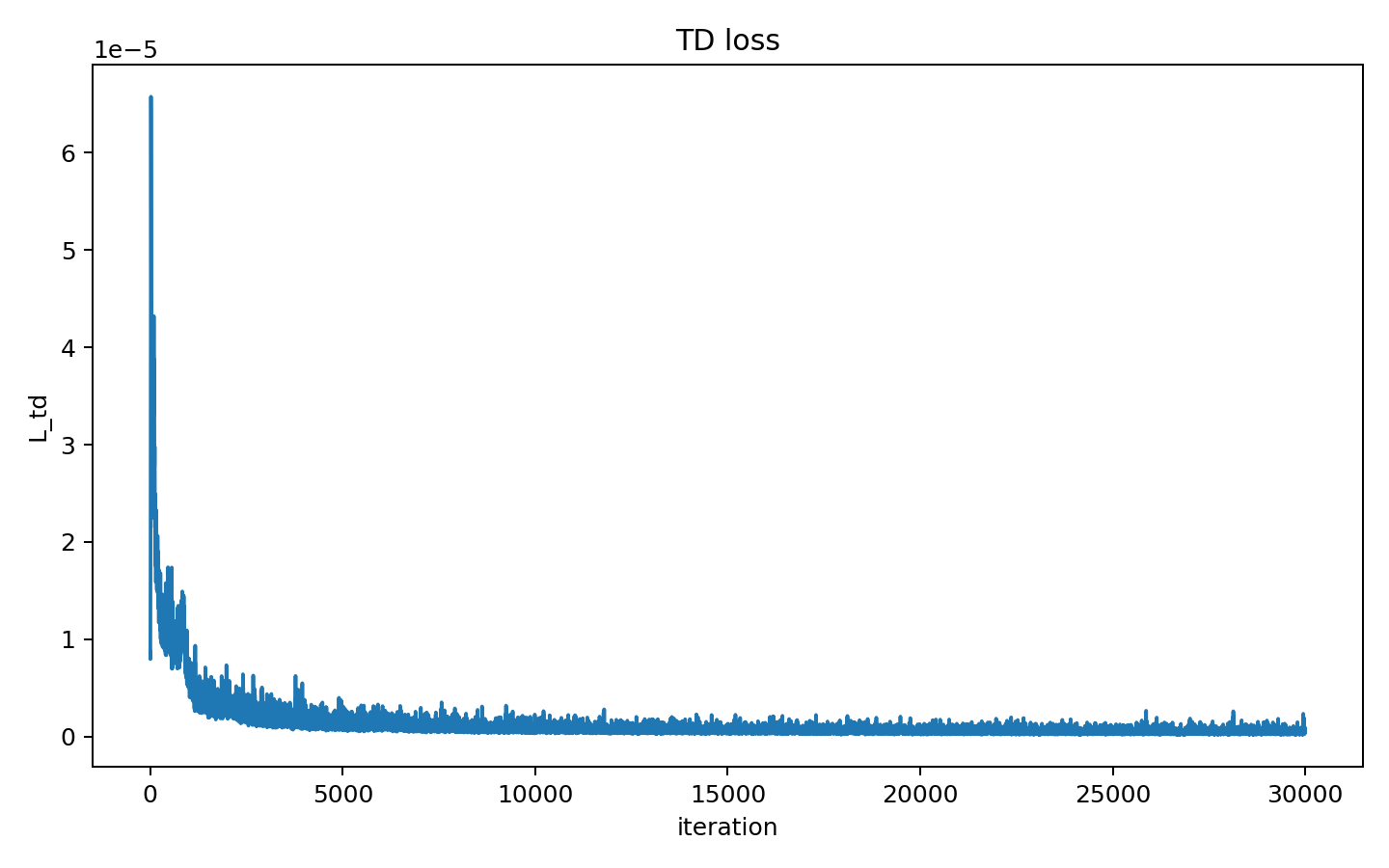}
    \caption{TD-loss trajectory for the 6D chained-integrator experiment.}
    \label{fig:td_6d}
\end{figure}

\begin{figure}[H]
    \centering
    \includegraphics[width=0.8\linewidth]{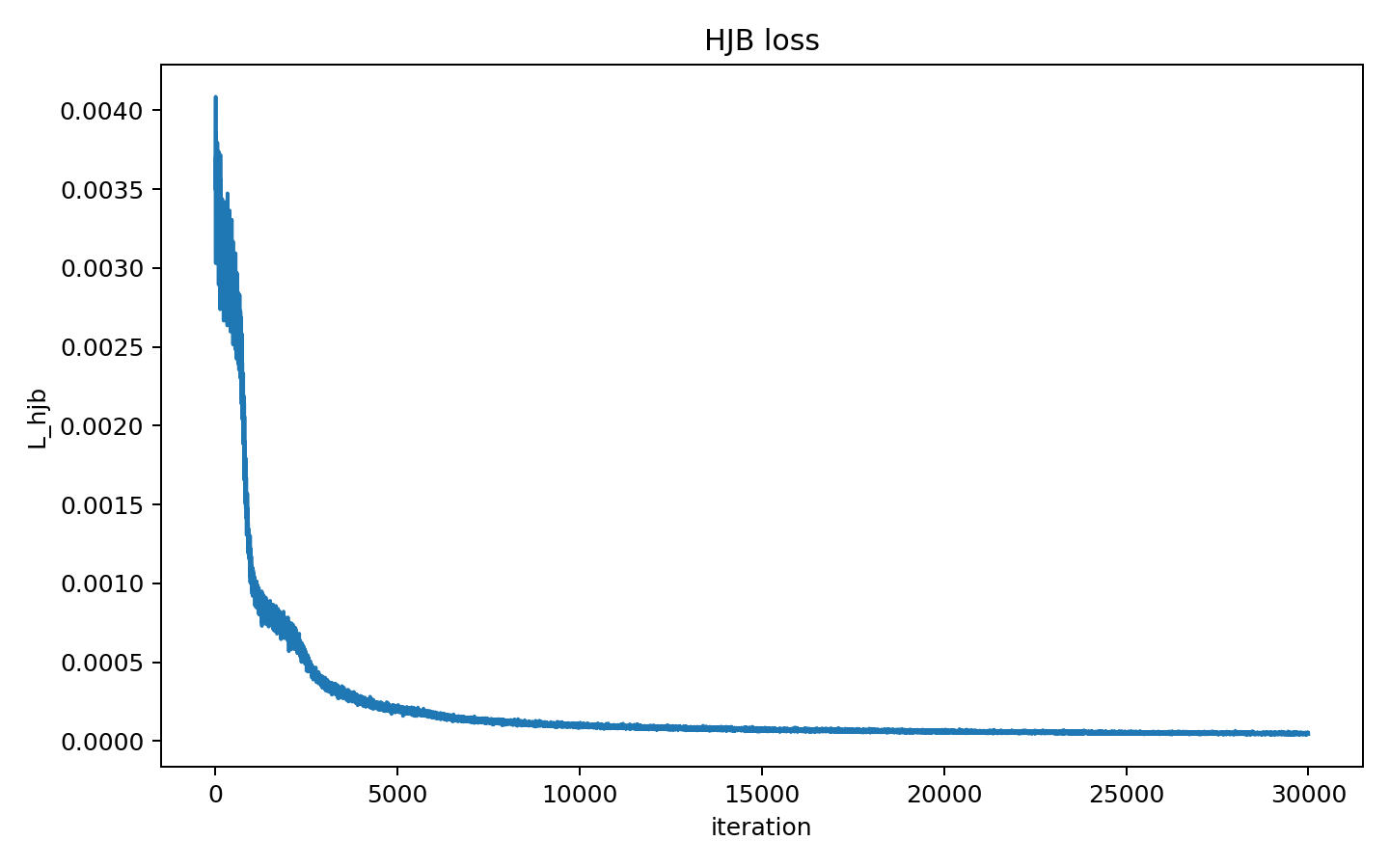}
    \caption{Numerically evaluated HJB loss during TD-only training for the 6D chained-integrator experiment.}
    \label{fig:hjb_6d}
\end{figure}

Figures~\ref{fig:td_4d}--\ref{fig:hjb_6d} show the resulting TD- and HJB-loss curves. In both systems, the TD loss decreases over training, and the numerically evaluated HJB loss also decreases, despite not being included in the optimization objective. Although these higher-dimensional experiments do not provide full-grid PDE comparisons as in the 2D case, they nevertheless provide additional empirical support for the main claim that TD training alone drives the learned value toward the HJB solution in higher dimensions.

\section{Reproducibility details for experiments.}
\label{app:reproducibility_details}

\begin{table}[H]
\centering
\caption{Reproducibility details for the 2D \(\lambda\)-sweep experiments.}
\label{tab:repro_2d_lambda}
\footnotesize
\setlength{\tabcolsep}{4pt}
\renewcommand{\arraystretch}{1.05}
\begin{tabularx}{\columnwidth}{@{}lX@{}}
\toprule
Setting & Value \\
\midrule
System & Double integrator \\
State & $(x_1,x_2)$ \\
Control & $u\in[-1,1]$ \\
Actions & $\{-1,+1\}$ \\
$\Delta\tau$ & 0.05 \\
$\lambda$ & $\{1.0,1.5,2.0,2.5,3.0\}$ \\
Training & TD only \\
Sampler & iid uniform \\
Discretization & Euler \\
Batch size & 8192 \\
Learning rate/Adam step & $3\times 10^{-4}$ \\
Hidden layers & $(256,256)$ \\
Initialization & SIREN ($w_0=30$) \\
Iterations & 2000 \\
ROI & $[-2.5,2.5]^2$ \\
Cost & Clipped travel cost, $r=0.5$, $\alpha=1$ \\
Monitor & TD loss, HJB loss \\
\bottomrule
\end{tabularx}
\end{table}

\begin{table}[H]
\centering
\caption{Reproducibility details for the 4D experiment.}
\label{tab:repro_4d}
\footnotesize
\setlength{\tabcolsep}{4pt}
\renewcommand{\arraystretch}{1.05}
\begin{tabularx}{\columnwidth}{@{}lX@{}}
\toprule
Setting & Value \\
\midrule
System & Planar double integrator \\
State & $(p_x,p_y,v_x,v_y)$ \\
Control & $(u_x,u_y)\in[-1,1]^2$ \\
Actions & $\{(-1,-1),(-1,1),(1,-1),(1,1)\}$ \\
$\Delta\tau$ & 0.05 \\
$\lambda$ & 1.0 \\
Training & TD only \\
Sampler & iid uniform \\
Discretization & Euler \\
Batch size & 8192 \\
Learning rate/Adam step & $3\times 10^{-4}$ \\
Hidden layers & $(256,256)$ \\
Initialization & SIREN ($w_0=30$) \\
Iterations & 10000 \\
ROI & $[-2.5,2.5]^4$ \\
Cost & Clipped travel cost, $r=0.5$, $\alpha=1$ \\
Monitor & TD loss, HJB loss \\
\bottomrule
\end{tabularx}
\end{table}

\begin{table}[H]
\centering
\caption{Reproducibility details for the 6D experiment.}
\label{tab:repro_6d}
\footnotesize
\setlength{\tabcolsep}{4pt}
\renewcommand{\arraystretch}{1.05}
\begin{tabularx}{\columnwidth}{@{}lX@{}}
\toprule
Setting & Value \\
\midrule
System & Chained integrator \\
State & $(p_x,p_y,v_x,v_y,a_x,a_y)$ \\
Control & $(j_x,j_y)\in[-1,1]^2$ \\
Actions & $\{-1,0,1\}^2$ \\
$\Delta\tau$ & 0.05 \\
$\lambda$ & 1.0 \\
Training & TD only \\
Sampler & iid uniform \\
Discretization & Euler \\
Batch size & 16384 \\
Learning rate/Adam step & $10^{-4}$ \\
Hidden layers & $(256,256,256)$ \\
Initialization & SIREN ($\omega_{0} = 30$) \\
Iterations & 30000 \\
ROI & $[-1.5,1.5]^4\times[-1.0,1.0]^2$ \\
Cost & Clipped travel cost, $r=0.5$, $\alpha=1$ \\
Monitor & TD loss, HJB loss \\
\bottomrule
\end{tabularx}
\end{table}


\input{}

\end{document}